\documentclass[conference]{IEEEtran}
\IEEEoverridecommandlockouts

\usepackage{cite}
\usepackage{amsmath,amssymb,amsfonts}
\usepackage[linesnumbered,ruled]{algorithm2e}
\usepackage{graphicx}
\usepackage{textcomp}
\usepackage{xcolor}
\usepackage{bm}
\usepackage{adjustbox}
\usepackage{tabularx,booktabs}
\usepackage{subcaption}
\usepackage{varwidth}
\usepackage{enumitem}
\usepackage{tikz}
\usepackage[
n,
advantage,
operators,
sets,
adversary,
landau,
probability,
notions,
logic,
ff,
mm,
primitives,
events,
complexity,
asymptotics,
keys
]{cryptocode}
\def\BibTeX{{\rm B\kern-.05em{\sc i\kern-.025em b}\kern-.08em
    T\kern-.1667em\lower.7ex\hbox{E}\kern-.125emX}}

\newcommand*{\xdash}[1][3em]{\rule[0.5ex]{#1}{\heavyrulewidth}}

\begin{document}

\title{\textsc{BlindTrust}: Oblivious Remote Attestation for Secure Service Function Chains}

\author{\IEEEauthorblockN{Heini Bergsson Debes$^{*}$, Thanassis Giannetsos$^{\dagger}$, Ioannis Krontiris$^{\mp}$}
\IEEEauthorblockA{Technical University of Denmark (DTU), Cyber Security Section, Denmark \\
	$^{\ddagger}$Ubitech Ltd., Digital Security \& Trusted Computing Group, Greece \\
	$^{\mp}$ European Research Center, Huawei Technologies, Munich, Germany \\
	Email: \ heib@dtu.dk, agiannetsos@ubitech.eu, ioannis.krontiris@huawei.com}
	}

\newcommand{\orc}{\mathcal{O}\ensuremath{\mathsf{rc}}}
\newcommand{\vf}{\mathcal{VF}}
\newcommand{\aagent}{\mathcal{AA}\ensuremath{\mathsf{gt}}}
\newcommand{\trace}{\mathcal{T}\ensuremath{\mathsf{rce}}}
\renewcommand{\verifier}{\mathcal{V}\ensuremath{\mathsf{rf}}}
\renewcommand{\prover}{\mathcal{P}\ensuremath{\mathsf{rv}}}
\newcommand{\tc}{\mathcal{TC}}

\maketitle

\newtheorem{definition}{Definition}
\newtheorem{property}{Property}

\begin{abstract}
With the rapidly evolving next-generation systems-of-systems, we face new security, resilience, and operational assurance challenges. In the face of the increasing attack landscape, it is necessary to cater to efficient mechanisms to verify software and device integrity to detect run-time modifications. Towards this direction, remote attestation is a promising defense mechanism that allows a third party, the verifier, to ensure a remote device's (the prover's) integrity. However, many of the existing families of attestation solutions have strong assumptions on the verifying entity's trustworthiness, thus not allowing for privacy-preserving integrity correctness. Furthermore, they suffer from scalability and efficiency issues. This paper presents a lightweight dynamic configuration integrity verification that enables inter and intra-device attestation without disclosing any configuration information and can be applied on both resource-constrained edge devices and cloud services. Our goal is to enhance run-time software integrity and trustworthiness with a scalable solution eliminating the need for federated infrastructure trust. 
\end{abstract}


\section{Introduction}
\label{sec:introduction}

Recently, academia and industry working groups have made substantial efforts towards realizing next-generation smart-connectivity ``\emph{Systems-of-Systems}'' (SoS). These systems have evolved from local, standalone systems into safe and secure solutions distributed over the continuum from cyber-physical end devices, to edge servers and cloud facilities. The core pillar in such ecosystems is the establishment of trust-aware Service Graph Chains (SGCs) comprising both resource-constrained devices, running at the edge, but also container-based technologies (e.g., Docker, LXC, rkt)~\cite{de2019integrity}. 


The primary existing mechanisms to establish trust is by leveraging the concept of trusted computing~\cite{sailer2004design, de2019integrity, luo2019container, chen2008property}, which addresses the need for verifiable evidence about a system and the integrity of its trusted computing base and, to this end, related specifications provide the foundational concepts such as \emph{measured boot} and \emph{remote attestation}. Within the realms of malware detection, remote attestation (RA) emerged as a simple challenge-response protocol to enable a verifier ($\verifier$) to ascertain the integrity of a remote platform, the prover ($\prover$). A key component in building such trusted computing systems is a highly secure anchor (either software- or hardware-based) that serves as a Root-of-Trust (RoT) towards providing cryptographic functions, measuring and reporting the behavior of running software, and storing data securely. Prominent examples include Trusted Execution Environments (e.g., TrustZone)~\cite{sabt2015trusted} and Trusted Platform Modules (TPMs)~\cite{tcgTpmArchitecture}.


However, none of them is sufficient to deal with the pressing challenge that container-based virtualization faces concerning assumptions on the trustworthiness of the $\verifier$ entity: it should be difficult for any (possibly compromised) $\verifier$ to infer any meaningful information on the state or configuration of any of the devices or containers comprising the service graph chain. In this context, it is essential to ensure not only the security of the underlying host and other loaded containers but also their privacy and confidentiality - \emph{an attacker should not be able to infer any information on the configuration of any other container loaded in the same containerized node or virtual function}.

This dictates for an \emph{oblivious} theme of building trust for such SoS where a $\prover$ can attest all of its components without the need to reveal specific configuration details of its software stack. For instance, suppose that a $\prover$ runs a Python interpreter. The $\prover$ may wish not to reveal that it runs version 2.7.13 of the CPython implementation. One option would be to introduce ambiguity about the software stack components (e.g., by having the $\prover$ only reveal that it has a CPython implementation), thus making it harder for a malicious $\verifier$ to exploit zero-day vulnerabilities in the $\prover$'s code directly. However, an even stronger claim is to have the $\prover$ not reveal \emph{anything}, which would make it \emph{impossible} for $\verifier$s to infer \emph{anything}. However, this sets the challenge ahead: \emph{How can a $\prover$ prove its integrity correctness without disclosing any information about its software stack's configuration?}

One overarching approach, which is the bedrock of the presented work, is to have a centralized entity (e.g., orchestrator in charge of deploying and managing the lifecycle of nodes) who determines what is correct and what is not, and then have that party setup appropriate cryptographic material (i.e., restrained attestation keys) on each node in the network and distribute them to all neighboring nodes. The ability to then use such restrained keys is physically ``locked'' from the node until the node can prove its correctness - supply correct measurements that will ``unlock'' its usage. Once released, the node can use the key to sign nonces supplied from the surrounding $\verifier$s, acting as verifiable statements about its state so that other components can align their actions appropriately and an overall system state can be accessed and verified. Similarly, if $\verifier$s receive no response or the signature is not produced using the key initially agreed upon and advertised by the centralized entity, they can justifiably assume that the $\prover$ is untrusted. Note that $\verifier$s need only to know that the $\prover$ is in an authorized state, not what that state is. However, one main challenge of such approaches is the strong link between the restrained cryptographic material and the specific Configuration Integrity Verification (CIV) policies: Whenever an updated policy must be enforced, due to a change to the configuration of the overall system, a new attestation key must be created~\cite{larsen2020cloudvaults}. Managing and updating such symmetric secrets creates an overwhelming \emph{key distribution problem}.

\textbf{\emph{Contributions:}} This paper provides a novel CIV protocol for supporting trust-aware SGCs with verifiable evidence on the integrity and correctness of deployed devices and virtual functions. Key features provided that extend the state-of-the-art include the: (\emph{i}) possibility to distinguish which container is compromised, and (\emph{ii}) the use of trusted computing for enabling inter- and intra-device attestation without disclosing any configuration information. Our proposed solution is scalable, (partially) decentralized, and capable of withstanding even a prolonged siege by a pre-determined attacker as the system can dynamically adapt to its security and trust state. We demonstrate our scheme with an implementation leveraging a Trusted Platform Module (TPM), following the TCG TPM 2.0 specification~\cite{tcgTpmArchitecture}, and benchmark its performance. 

\section{Background and Related Work}
\label{sec:backgroundAndRelatedWork}

\subsection{Preliminary Definitions}

\subsubsection{Building Chains of Trust with Monotonic Counters}
\label{subsubsec:monotonicCounters}

To enforce secure boot on machine $m$, we can require that all components verify their successors by the following recurrence construct: $I_0 = \true;\allowbreak I_{i+1} = I_i \land V_i(L_{i+1})$, where $I_i$ denotes the integrity of layer $i$ and $V_i$ is the corresponding verification function which compares the hash of its successor with a trusted reference value (TRV). If verification fails at any layer, the lower layer refuses to pass control and bricks the boot process. However, to relax the boot process, we can hold off verification and instead have the components record their successors' measurements. To facilitate such recording, each TPM has several PCRs that can only be modified in two ways: (\emph{i}) by resetting the machine on which the TPM resides, and (\emph{ii}) through a interface called \texttt{PCR\_Extend}, which takes a value $v$ and a PCR $i$ as arguments and then aggregates $v$ and the existing PCR value $PCR_i$ by computing: $PCR_i \gets \hash(PCR_i \, \concat \, \hash(v))$. The irreversibility property of PCRs makes it possible to build strong chains of trust. In the context of measured boot, if all components in the boot sequence: $\langle init,\allowbreak BL(m),\allowbreak OS(m),\allowbreak APP(m) \rangle$ are measured into $PCR_j$, where $init$ is the initial value that $PCR_j$ is reset to and $v_1, \ldots, v_n$ are the corresponding TRVs of $m$'s components, then the $PCR_j$ aggregate corresponds to a trusted boot if $PCR_j =\allowbreak \hash(\allowbreak\ldots\allowbreak(\allowbreak\hash(\allowbreak init\allowbreak\, \concat \, \hash(v_1)) \, \concat \, \hash(v_2))\ldots \, \concat \, \hash(v_n))$.

\subsubsection{Remote Attestation}
\label{subsubsec:ra}

In the context of TPMs, we can use the \texttt{Quote} interface to get a signed report of select PCR aggregates. Thus, considering the example of measured boot in Section~\ref{subsubsec:monotonicCounters}, by presenting a \texttt{Quote}, we can delegate the verification of a machine's ($\prover$) boot process to a remote $\verifier$. The two most fundamental ways to run the RA protocol are:
\begin{enumerate}
    \item \textbf{Init:} $\verifier$ knows $TRV = \hash(\ldots(\hash(init \concat \hash(v_1))\allowbreak \concat \hash(v_2))\allowbreak\ldots\allowbreak \concat \hash(v_n))$.\\
    \textbf{Step 1:} $\prover$ sends $PCR_j$ to $\verifier$.\\ \textbf{Step 2:} $\mathsf{trusted}(\prover) \iff PCR_j = TRV$.
    \item \textbf{Init:} $\verifier$ knows $init, TRV = \set{v_1, \ldots, v_n}$.\\
    \textbf{Step 1:} $\prover$ sends $PCR_j, L = \langle v_1^\prime, \ldots, v_n^\prime \rangle$ to $\verifier$.\\
    \textbf{Step 2:} $\mathsf{trusted}(\prover) \iff PCR_j = \hash(\ldots(\hash(init\allowbreak \concat\allowbreak \hash(\allowbreak L_1\allowbreak))\allowbreak \concat \hash(L_2))\allowbreak\ldots\allowbreak \concat \hash(L_n)) \land \forall v^\prime \in L: v^\prime \in TRV$.
\end{enumerate}

In the first setup, $\verifier$ only knows a TRV of the trusted boot process. However, if the measurement chain continues beyond the boot process, or the order in which components are loaded is non-deterministic, having a single TRV is insufficient. For non-deterministic temporal orders (e.g., during run-time), it is preferred to keep a log $L$ to record the measurements' order. Thus, in the latter setup, when $\verifier$, who has a list of TRVs, wants to determine $\prover$'s state, $\prover$ sends $L$ and a \texttt{Quote} over $PCR_j$ to $\verifier$ who: (\emph{i}) validates the association between $PCR_j$ and $L$ by re-creating the aggregate from $L$'s entries and comparing it to $PCR_j$, and (\emph{ii}) compares all of $L$'s entries to its TRV list. If everything holds, then $\prover$ is in a trusted state.




\subsection{Toward Confidential Configuration Integrity Verification}
\label{sec:relatedWork}

IMA~\cite{sailer2004design} is the backbone of several schemes centered in measuring container integrity~\cite{de2019integrity, luo2019container}. It extends measured boot into the OS, where, depending on a measurement policy (MP), files and binaries (objects) are measured and recorded in a measurement log (ML) and a TPM register. Depending on MP, IMA proceeds to continuously remeasure objects as they are accessed or changed during run-time. However, since IMA assumes the second setup of Section~\ref{subsubsec:ra}, RA is impractical in large networks where all participants must diligently maintain an excessive list of TRVs. Further, since the protocol requires that ML and the quoted information be sent to a $\verifier$, it exhibits configuration confidentiality issues: (\emph{i}) if $\prover$ has sensitive objects, they too must be admitted for $\verifier$ to determine their correctness; (\emph{ii}) if $\prover$ records multiple containers in the same ML and PCR, $\verifier$ learns about each container; (\emph{iii}) if $\verifier$ is dishonest, she benefits from ML since she can identify and spear-phish vulnerable components.

Since IMA's default ML template contains few associators, DIVE~\cite{de2019integrity} introduce a \texttt{dev-id} to link entries with containers. Thus, if $\verifier$ wants to ascertain container $c$'s correctness, only $c$'s ML entries need to be verified against TRVs. However, since $\verifier$ learns the full ML, excessive configuration exposure remains an issue. Solving the ML multiplexing issue, security namespaces~\cite{sun2018security} enable segregating containers such that containers have separate MLs and PCRs. However, associating unique PCRs to containers only works so long as there are fewer containers than PCRs. Further, although $\prover$ only sends one container's ML and PCR aggregate per request, nothing stops $\verifier$ from querying all containers. To mitigate the issue, Container-IMA~\cite{luo2019container} assume a \emph{secret} between kernel space and the participant that spawned a container $c$. When $\verifier$ queries $c$, $\prover$ sends $c$'s measurements obscured under $c$'s secret, thus preventing exposure to $\verifier$s unaware of $c$'s secret. The main deficiency, however, is that only $c$'s parent can verify $c$.

While there exist other RA variants, e.g., Property-Based Attestation (PBA)~\cite{chen2008property}, where many measurements are mapped to one property to prevent $\verifier$ from learning $\prover$'s exact configuration, the overhead of requiring that participants agree on TRVs or what constitutes ``property fulfillment'' remains an issue. To mitigate the issue, CloudVaults~\cite{larsen2020cloudvaults} proposed a scheme wherein a system orchestrator ($\orc$), who knows each participant's TRVs, securely establishes TRV-constrained asymmetric attestation key (AK) pairs in each participant's TPM, where the secret AK can be used only if a specified PCR contains the TRV. Thus, $\verifier$s that know a $\prover$'s public AK can send $\prover$ a fresh challenge, and if $\prover$ replies with a signature over the challenge using its secret AK, then $\verifier$ knows that $\prover$ is in a correct state. The problem, however, is that new AKs must be created and shared with all participants whenever configurations change, causing a key-distribution problem. 






\section{Toward Oblivious Remote Attestation}

\subsection{Notation}

We consider the following symbols and abbreviations:

{\small
\begin{description}[leftmargin=!,labelwidth=\widthof{$RC(\eval(\mathsf{cmd}))$},align=right]
\item[$\vf$] A virtual function.
\item[$\tc$] Trusted Component (e.g., a SW or HW-TPM).
\item[$vTPM$] A virtual (softwarized) Trusted Platform Module.
\item[$\orc$] The orchestrator (trusted authority).
\item[$\aagent^\mathcal{X}$] Local attestation agent running on $\vf$ $\mathcal{X}$.
\item[$\trace(r)$] Retrieve the binary contents of object identified by $r$ using the secure and immutable tracer $\trace$.
\item[$\gets, =$] $\gets$ denotes assignment and $=$ denotes comparison.
\item[$h$] Hash digest ($0\ldots0$ is used to denote a zero-digest).
\item[$\hash$] A secure and collision-resistant hash function.
\item[$\hk_{(A, B)}$] Symmetric hash key known only by $A$ and $B$.
\item[$\mathsf{HMAC}(\hk, i)$] $\hk$-keyed Message Authentication Code over $i$.
\item[$\eval(expr)$] Evaluation function for arbitrary expressions $expr$.
\item[$\verify(expr)$] Verification, which interrupts if $\eval(expr) = 0$.
\item[$\sign(m, \key)$] Computes a signature over $m$ using $\key$.
\item[$\sig^\key_\phi$] Signature over $\phi$ using key $\key$.
\item[$\mathcal{H}$] TPM handle, where $\mathcal{H} \in \NN$.
\item[$\mathsf{TPL}(\phi)$] Template for object $\phi$ (including its attributes).
\item[$\mathcal{B}$] Boolean variable: $\mathcal{B} \in \mathbb{B} = \set{0, 1} = \set{\false, \true}$.
\item[$\mathsf{name}(\phi)$] $\phi$'s name. For keys and NV indices, it is a digest over the public area, including attributes and policy.
\item[$CC_{\mathsf{cmd}}$] $\mathsf{cmd}$'s TPM Command Code.
\item[$RC(\eval(\mathsf{cmd}))$] TPM Response Code after executing $\mathsf{cmd}$.
\item[$mPCR^{\vf}$] Set of mock PCR tuples: $\{\langle idx_0, h_0 \rangle, \ldots, \langle idx_n,$ $h_n,\rangle\}$ associated with $\vf$, where $idx_i \in \NN_0$.
\item[$mNVPCR^{\vf}$] Set of mock NV PCR tuples: $\{\langle \mathcal{H}_0, h_0, \mathsf{name}(\mathcal{H}_0)$ $\rangle, \ldots, \langle \mathcal{H}_n, h_n, \mathsf{name}(\mathcal{H}_n) \rangle\}$ associated with $\vf$.
\item[$PCRS$] Set of PCR selectors: $\set{i\, :\, i \in \NN_0}$.
\item[$NVPCRS$] Set of NV PCR selector tuples: $\{\langle \mathcal{H}_0, h_0 \rangle, \ldots, \langle$ $\mathcal{H}_n, h_n \rangle\}$.
\item[$PPS$] A TPM's \emph{secret} Platform Primary Seed.
\item[$\mathsf{proof}(\phi)$] A TPM's \emph{secret} value associated $\phi$'s hierarchy.
\item[$SK$] Restricted storage (decryption) key.
\item[$EK^\mathcal{O}_p$] $\mathcal{O}$'s endorsement (restricted signing) key pair: $\langle$ $EK_\pk, EK_\sk \rangle$, where $EK_\sk$ is encrypted, denoted $\mathsf{sealed}(EK_\sk)$, while outside the TPM. Optionally, $p$ is used to refer to a specific part of the EK.
\item[$AK^\mathcal{O}_p$] $\mathcal{O}$'s attestation (unrestricted signing) key pair: $\langle AK_\pk, AK_\sk \rangle$, where $AK_\sk$ is encrypted, denoted $\mathsf{sealed}(AK_\sk)$, while outside the TPM. Optionally, $p$ is used to refer to a specific part of the AK.
\end{description}
}\normalsize

\subsection{System and Threat Model}

\subsubsection{System Model}
\label{subsubsec:systemModel}

\begin{figure}[htbp]
    \centering
    \includegraphics[width=1\columnwidth]{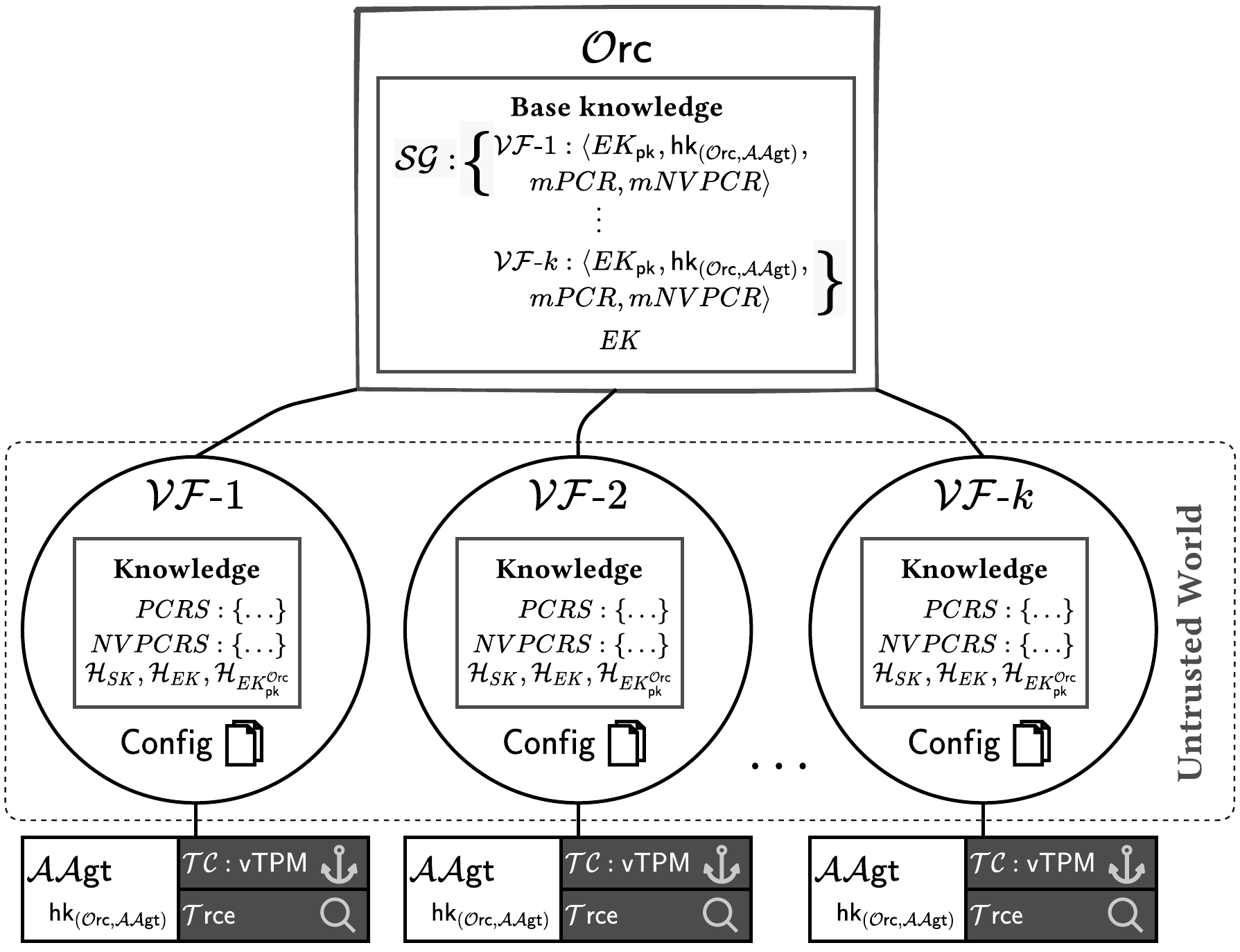}
    \caption{Conceptual (initial) system knowledge model.}
    \label{fig:systemModel}
    \vspace{-1em}
\end{figure}

The considered system (Fig.~\ref{fig:systemModel}) is composed of a virtualized network infrastructure where an orchestrator ($\orc$) spawns and governs a set of heterogeneous and cloud-native containerized $\vf$ instances as part of dedicated Service Graph ($\mathcal{SG}$) chains. Each deployed $\vf$ is associated with three $\tc$s: a vTPM, serving as its trust anchor, an attestation agent ($\aagent$) to service inquiries, and a secure tracer ($\trace$) to measure the current state of a $\vf$'s configuration (Definition~\ref{def:config}), ranging from its base software image, platform-specific information, and other binaries. Whether the vTPM is anchored to an HW-TPM~\cite{perez2006vtpm} to provide enhanced security guarantees is a design choice and is beyond this paper's scope.

\normalsize

\begin{definition}[Config]\label{def:config}
A $\vf$'s configuration set represents all of its uniquely identifiable objects (blobs of binary data).
\end{definition}

To proactively secure a $\vf$'s participation in the $\mathcal{SG}$, $\orc$ intermittently demands a $\vf$ to re-measure parts (or all) of its configuration into its vTPM's PCRs (either normal or NV-based) to justify its conformance with the currently compulsory policies. To track active PCRs, $\vf$s maintain a separate list for normal (PCRS) and NV-based PCRs (NVPCRS). Each $\vf$ also begins with three persistent vTPM key handles: (\emph{i}) a vTPM storage key (SK) to enable the creation of AKs, (\emph{ii}) the $\vf$'s unique EK, which was agreed upon with $\orc$ during deployment, and (\emph{iii}) the public part of $\orc$'s EK to authenticate $\orc$. Further, we assume a secret symmetric hash key ($\hk$) shared between $\orc$ and each $\aagent$ to enable $\aagent$s to authenticate their involvement in measurements. The hash key is assumed to reside in secure storage, inaccessible to any software, except for privileged code of the local $\aagent$. 

On $\orc$, each $\vf$ (besides its identity) is initially represented by the certified public part of its EK, the hash key shared with the $\vf$'s $\aagent$, and two sets of mock PCRs, one representing the mock (emulated) state the $\vf$'s normal PCRs (mPCR), and another of the NV-based PCRs (mNVPCR).

\subsubsection{Adversarial Model}
\label{subsec:adversarialModel}

We consider configuration integrity and therefore do not consider stateless attacks where $\adv$ performs nefarious tasks without touching any configuration (by Definition~\ref{def:config}). We assume that the underlying system maintains appropriate file metadata structures for each identifiable object in a $\vf$'s configuration and cannot be altered by $\adv$. Metadata that relate to the object's integrity (e.g., its creation and modification timestamps, or \texttt{i\_generation} and \texttt{i\_version} for Linux kernel's mounted with \emph{inode} versioning support) are assumed to be included in an object's measurements to prevent $\adv$ from unnoticeably recording a $\vf$'s configuration, alter it, and then restore it before the $\vf$ is told by $\orc$ to re-measure its configuration. The untrusted zone, where our protocol is designed to secure, is depicted in Fig.~\ref{fig:systemModel}. We let our adversary ($\adv$) roam freely in the untrusted zone (a $\vf$'s userspace) with unrestricted (create, read, write, and delete) access, including oracle access to the attached $\tc$s. For incoming and outgoing messages, we restrict $\adv$ to the classical Dolev-Yao model, where $\adv$ cannot break cryptographic primitives but is free to intercept, block, replay, spoof, and inject messages on the channel from any source. Thus, besides its local knowledge, unless $\adv$ learns new cryptographic keys from participating in the protocol or deriving them as part of other messages, she cannot compose messages using the secret keys of other participants. As a final note, we assume that unresponsive $\vf$s (within reasonable bounds) are untrusted, which, when noticed by $\orc$, triggers the revocation of its AK throughout $\mathcal{SG}$.




\subsection{High-Level Security Properties}
\label{subsec:highLevelSecurityProperties}

The objective of our protocol is twofold: (\emph{i}) to enable $\orc$ to securely enroll $\vf$s in the $\mathcal{SG}$, and (\emph{ii}) to enable enrolled $\vf$s to perform configuration-oblivious inter-$\vf$ CIV. Specifically, our scheme is designed to provide the following properties:

\begin{property}[Configuration Correctness]\label{property:configurationCorrectness}
A $\vf$'s load-time and run-time configurations (by Definition~\ref{def:config}) must have adhered to the latest attestation policy authorized by $\orc$, in order to be verified as being correct by any other $\vf$s.
\end{property}

\begin{property}[Secure Enrollment]\label{property:remoteDeployment}
To guard the attestation-enhanced division of the $\mathcal{SG}$, a $\vf$'s enrollment involves $\orc$ supervising the $\vf$ in creating an acceptable Attestation Key (AK), which is certified to remain under $\orc$'s control.
\end{property}

\begin{property}[Forward Acceptance]\label{property:forwardAcceptance}
To prevent excessive AK recreation and redistribution, all AKs are created such that they can be continuously repurposed (i.e., in which policy they attest) as determined and authorized by $\orc$, thus keeping key distribution at a minimum and circumventing the performance cost of creating and redistributing multiple AKs for each $\vf$.
\end{property}

\begin{property}[Freshness]\label{property:policyFreshness}
To ensure non-ambiguous verification, a $\vf$ can have at most one policy that unlocks its AK.
\end{property}


\begin{property}[Zero-Knowledge CIV]\label{prop:zeroKnowledgeVerification}
To keep configurations confidential, any $\vf$ should only require another $\vf^\prime$'s AK's public part ($AK^{\vf^\prime}_{\pk}$) to verify its configuration correctness.
\end{property}

Note that in our considered setup (Section~\ref{subsubsec:systemModel}), $\orc$ is considered the central entity (i.e., policy creator, authorizer, and enforcer) who knows the configuration of each $\vf$ since it creates all $\vf$s and manages the whole lifecycle of $\mathcal{SG}$ chains. However, for the protocol to work, $\orc$ is \emph{only} required to be online during a $\vf$'s enrollment and when new configurations need to be deployed. Once enrolled, $\vf$s run the remaining protocol among themselves in a decentralized manner.



\section{An Architectural Blueprint}

\subsection{High-Level Overview}

\begin{figure}[htbp]
    \centering
    \includegraphics[width=1\columnwidth]{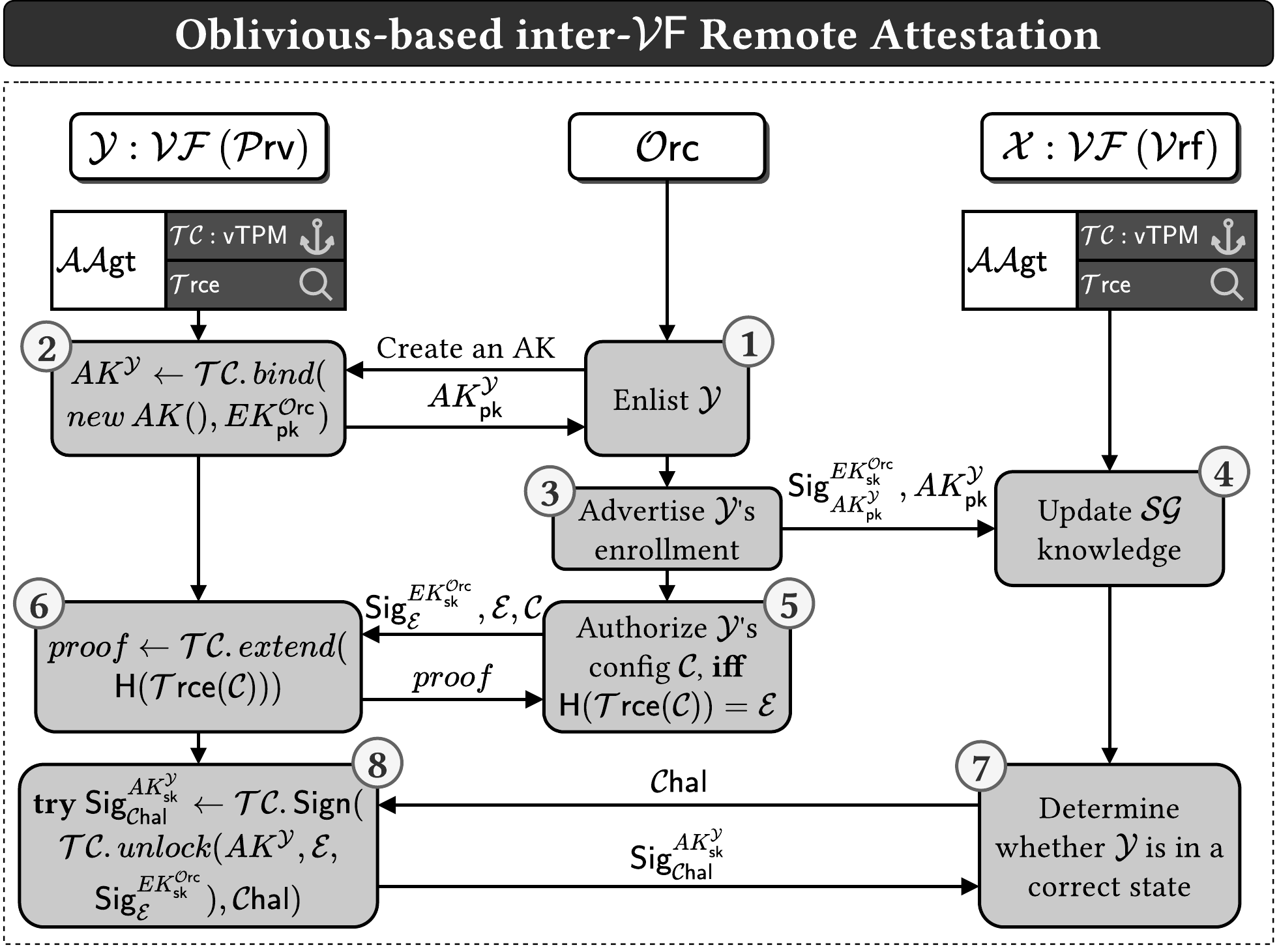}
    \caption{Holistic work-flow of the ORA protocol.}
    \label{fig:workFlow}
    \vspace{-1em}
\end{figure}

By conditioning a $\vf$'s ability to attest on whether its configurations are authorized by $\orc$, the ORA scheme (see Fig.~\ref{fig:workFlow}) enables arbitrary $\vf$s to verify the integrity of other $\vf$s while remaining oblivious to what constitutes their state. We preserve privacy as no exchange of platform or state details is required among $\vf$s. Specifically, contrary to using TPM \texttt{Quotes}, $\vf$s need no reference values to verify other $\vf$s.


\normalsize

The scheme's work-flow (Fig.~\ref{fig:workFlow}) is as follows. Let $\mathcal{SG} = \set{\mathcal{X} : \langle \ldots \rangle, \ldots}$ be the $\mathcal{SG}$ maintained by $\orc$. When a new $\vf$, say $\mathcal{Y}$, wishes to join, $\orc$ requests it to first create an AK (Step 1), $AK^\mathcal{Y}$, using its vTPM, and lock it to a \emph{flexible policy} bound to $\orc$'s EK, ensuring that only $\orc$ can permit $AK^\mathcal{Y}$'s use (Step 2). Once $AK^\mathcal{Y}$ is created, and $\orc$ has verified that it was done correctly, $\orc$ certifies $AK^\mathcal{Y}$ and advertises $\mathcal{Y}$'s enrollment to the appropriate $\mathcal{SG}$ chain (Steps 3), where existing $\vf$'s will include $\mathcal{Y}$ as an eligible peer (Step 4). Then, to enable $\mathcal{Y}$ to prove its configuration correctness using its AK, $\orc$ authorizes (signs using $EK^{\orc}_{\sk}$) a policy digest $\mathcal{E}$ over $\mathcal{Y}$'s currently acceptable configuration state, and sends it to $\mathcal{Y}$ (Step 5). Given the update request, $\mathcal{Y}$ measures its actual configuration into its vTPM (Step 6). When another $\vf$, $\mathcal{X}$, in the same $\mathcal{SG}$ chain as $\mathcal{Y}$, wants to determine whether $\mathcal{Y}$ is in a trusted state, it sends a challenge $\mathcal{C}\mathsf{hal}$ (e.g., a nonce) to $\mathcal{Y}$ (Step 7). If, and only if, $\mathcal{Y}$'s configuration measurements corresponded to what $\orc$ authorized, access is granted to use $AK^\mathcal{Y}$ to sign $\mathcal{C}\mathsf{hal}$ (Step 8). Note that steps 5 and 6 can repeat any number of times to change $\mathcal{Y}$'s trusted configuration state.


\subsection{Building Blocks}

Let us proceed with more details on the separate stages.

\subsubsection{AK Provisioning}
\label{subsubsec:akCreation}

\begin{figure}[htbp]
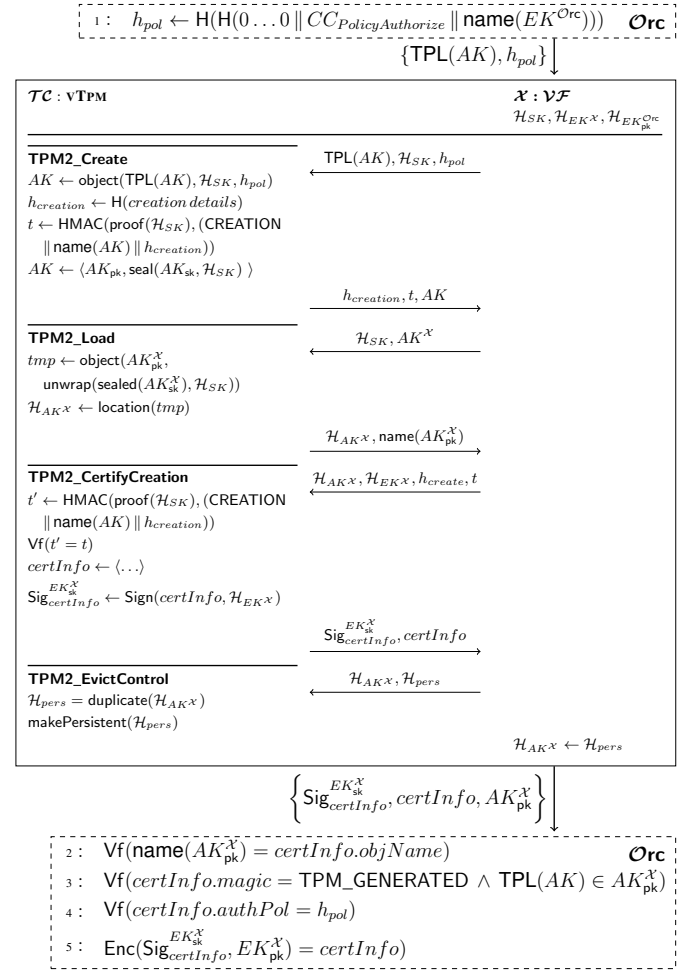

    \centering
    \begin{adjustbox}{
    , width=1\columnwidth}
    \begin{bbrenv}{C}
    
    \begin{bbrbox}[abovesep=0pt]
    \pseudocode[mode=text, colsep=0pt, colspace=0pt, jot=0mm]{
    $\bm{\mathcal{TC}} : \textsc{\textbf{vTpm}}$ \>\> $\bm{\mathcal{X}: \vf}$\\
    \>\> $\mathcal{H}_{SK}, \mathcal{H}_{EK^\mathcal{X}}, \mathcal{H}_{EK^{\orc}_\pk}$ \\[][\bottomrule] \\[-0.5\baselineskip]
    \> \sendmessage{<-}{style={line width=\lightrulewidth}, top={$\textsf{TPL}(AK), \mathcal{H}_{SK}, h_{pol}$}} \> \\[-2.2\baselineskip]
    \xdash[17.5em] \>\> \\[-0.5\baselineskip]
    $\bm{\mathsf{TPM2\_Create}}$ \>\> \\
    $AK \gets \mathsf{object}(\textsf{TPL}(AK), \mathcal{H}_{SK}, h_{pol})$ \>\> \\
    $h_{creation} \gets \hash(creation\,details)$ \>\> \\
    $t \gets \mathsf{HMAC}(\mathsf{proof}(\mathcal{H}_{SK}), (\mathsf{CREATION}$ \>\> \\
    $\quad \concat \, \textsf{name}(AK) \, \concat \, h_{creation}))$ \>\> \\
    $AK \gets \langle AK_\pk, \mathsf{seal}(AK_\sk, \mathcal{H}_{SK}) \ \rangle$ \>\> \\
    \> \sendmessage{->}{style={line width=\lightrulewidth}, top={$h_{creation}, t, AK$}} \> \\
    \> \sendmessage{<-}{style={line width=\lightrulewidth}, top={$\mathcal{H}_{SK}, AK^\mathcal{X}$}} \> \\[-2.2\baselineskip]
    \xdash[17.5em] \>\> \\[-0.5\baselineskip]
    $\bm{\mathsf{TPM2\_Load}}$ \>\> \\
    $tmp \gets \mathsf{object}(AK^\mathcal{X}_\pk, $ \>\> \\
    $\quad \mathsf{unwrap}(\mathsf{sealed}(AK^\mathcal{X}_\sk), \mathcal{H}_{SK}))$ \>\> \\
    $\mathcal{H}_{AK^\mathcal{X}} \gets \mathsf{location}(tmp)$ \>\> \\
    \> \sendmessage{->}{style={line width=\lightrulewidth}, top={$\mathcal{H}_{AK^\mathcal{X}}, \mathsf{name}(AK^\mathcal{X}_\pk)$}} \> \\
    \> \sendmessage{<-}{style={line width=\lightrulewidth}, top={$\mathcal{H}_{AK^\mathcal{X}}, \mathcal{H}_{EK^\mathcal{X}}, h_{create}, t$}} \> \\[-2.2\baselineskip]
    \xdash[17.5em] \>\> \\[-0.5\baselineskip]
    $\bm{\mathsf{TPM2\_CertifyCreation}}$ \>\> \\
    $t^\prime \gets \mathsf{HMAC}(\mathsf{proof}(\mathcal{H}_{SK}), (\mathsf{CREATION}$ \>\> \\
    $\quad \concat \, \textsf{name}(AK) \, \concat \, h_{creation}))$ \>\> \\
    $\verify(t^\prime = t)$ \>\> \\
    $certInfo \gets \langle \ldots \rangle$ \>\> \\
    $\sig^{EK^\mathcal{X}_\sk}_{certInfo} \gets \sign(certInfo, \mathcal{H}_{EK^\mathcal{X}})$ \>\> \\
    \> \sendmessage{->}{style={line width=\lightrulewidth}, top={$\sig^{EK^\mathcal{X}_\sk}_{certInfo}, certInfo$}} \> \\
    \> \sendmessage{<-}{style={line width=\lightrulewidth}, top={$\mathcal{H}_{AK^\mathcal{X}}, \mathcal{H}_{pers}$}} \> \\[-2.2\baselineskip]
    \xdash[17.5em] \>\> \\[-0.5\baselineskip]
    $\bm{\mathsf{TPM2\_EvictControl}}$ \>\> \\
    $\mathcal{H}_{pers} = \mathsf{duplicate}(\mathcal{H}_{AK^\mathcal{X}})$ \>\> \\
    $\mathsf{makePersistent}(\mathcal{H}_{pers})$ \>\> \\
    \>\> $\mathcal{H}_{AK^\mathcal{X}} \gets \mathcal{H}_{pers}$
    }
    \end{bbrbox}

    \node[draw, dashed, line width=\lightrulewidth, above=1.2cm of C.north east, anchor=east] (orc1) {\pseudocode[mode=text, codesize=\large, linenumbering]{
        $h_{pol} \gets \hash(\hash(0\ldots0 \, \concat \, CC_{PolicyAuthorize} \, \concat \, \textsf{name}(EK^{\orc})))\quad\quad\quad$
    }};
    \node[anchor=north east, xshift=-0.1cm, yshift=-0.1cm] at (orc1.north east) {\large{$\bm{\orc}$}};
    \draw[->, line width=\heavyrulewidth] ([xshift=-2.5cm]orc1.south east)--([xshift=-2.5cm, yshift=0.1cm]C.north east) node [midway, left] {\large $\set{\textsf{TPL}(AK), h_{pol}}$};
    
    \node[draw, dashed, line width=\lightrulewidth, below=2.8cm of C.south east, anchor=east] (orc2) {\pseudocode[mode=text, codesize=\large, linenumbering, lnstart=1]{
        $\verify(\textsf{name}(AK^\mathcal{X}_\pk) = certInfo.objName)$ \\
        $\verify(certInfo.magic = \mathsf{TPM\_GENERATED} \, \land \, \textsf{TPL}(AK) \in AK^\mathcal{X}_\pk)$\\
        $\verify(certInfo.authPol = h_{pol})$ \\
        $\enc(\sig^{EK^\mathcal{X}_\sk}_{certInfo}, EK^\mathcal{X}_\pk) = certInfo)$
    }};
    \node[anchor=north east, xshift=-0.1cm, yshift=-0.1cm] at (orc2.north east) {\large{$\bm{\orc}$}};
    \draw[->, line width=\heavyrulewidth] ([xshift=-2.5cm]C.south east)--([xshift=-2.5cm, yshift=0.1cm]orc2.north east) node [midway, left] {\large $\set{\sig^{EK^\mathcal{X}_\sk}_{certInfo}, certInfo, AK^\mathcal{X}_\pk}$};
    
    \end{bbrenv}
    \end{adjustbox}
    \caption{AK creation}
    \label{fig:akcreation}
\end{figure}

Fig.~\ref{fig:akcreation} shows the exchange of messages between the different actors in the AK-creation protocol, where $\orc$ is portrayed as an oracle who supplies input to and verifies output from the $\vf$ ($\mathcal{X}$). The protocol begins locally on $\orc$, where a policy digest is computed over the Command Code (CC) of \texttt{PolicyAuthorize} (specified in the specification~\cite{tcgTpmArchitecture}) and the name of $\orc$'s EK. Note that such policies are called \emph{flexible} since any object $\phi$ bound to the policy can only be used in a policy session with the vTPM after fulfilling some policy (e.g., that the PCRs are in a particular state) which the policy's owner ($\orc$ in our case) has authorized (signed). The policy digest, together with a template describing the key's characteristic traits (e.g., attributes and type), is then sent to $\mathcal{X}$, who forges the AK within its vTPM. Besides producing and returning the AK object, the vTPM also returns a signed ticket over the object to denote that it was created inside the vTPM. This ``creation'' ticket, together with the newly created AK object and $\mathcal{X}$'s EK, are then passed to \texttt{CertifyCreation}, where the vTPM vouches that it was involved in producing AK (if the ticket holds) by signing (using the supplied EK) the AK object along with some internal state information. Then, due to AK's flexibility, where AK can remain the same throughout $\mathcal{X}$'s lifetime, it is stored persistently in vTPM NV memory (using \texttt{EvictControl}). Finally, $\mathcal{X}$ presents the AK and certificate to $\orc$, who verifies the certificate's signature and scrutinizes its details to ascertain that the AK was created legitimately. If everything holds, $\mathcal{X}$ is permitted to participate in the $\mathcal{SG}$.



\normalsize

\subsubsection{Remote PCR Administration}
\label{subsubsec:remotePcrAdministration}


Although normal (static) PCRs cannot be reset during run-time, an NV slot defined to imitate a PCR can be deleted and recreated depending on how it is created. We, therefore, require that NV PCRs be created with a flexible policy, similar to AKs, such that \emph{only} upon deletion requests authorized by $\orc$ can the NV index be undefined. To ensure that only policies specifically authorized to undefine the NV index can be used, we additionally include the CC of \texttt{NV\_UndefineSpaceSpecial}, which requires that the $\orc$-signed policy bears a reference to \texttt{NV\_UndefineSpaceSpecial}. Further, to prevent a $\vf$ from undefining arbitrary NV indices, $\orc$ embeds into the policy a Command Parameter (CP) digest over the name of the NV index that should be undefined, which restricts the use of the policy only to be used on the correct NV index. Note that for brevity, the protocols to allocate and deallocate PCRs are given in Fig.~\ref{fig:createNvPcr} and Fig.~\ref{fig:pcrRevocation} of Appendix~\ref{appendix:protocols}, respectively, where we also elaborate more on the details of the processes. The important thing to note is that when any PCR (regular or NV) is attached to a $\vf$ $\mathcal{X}$, the new PCR index is added to $\mathcal{X}$'s local knowledge (its $PCRS$ and $NVPCRS$ structures), and also to $\orc$'s mock structures associated with $\mathcal{X}$, i.e., $mPCR^\mathcal{X}$ and $mNVPCR^\mathcal{X}$. By synchronizing active PCRs, a $\vf$ keeps an updated list of PCRs to attest. If the list is out of sync (or altered), attestation using its certified AK is futile.

\subsubsection{Supervised Updates}
\label{subsubsec:supervisedUpdates}

\begin{figure}[htbp]
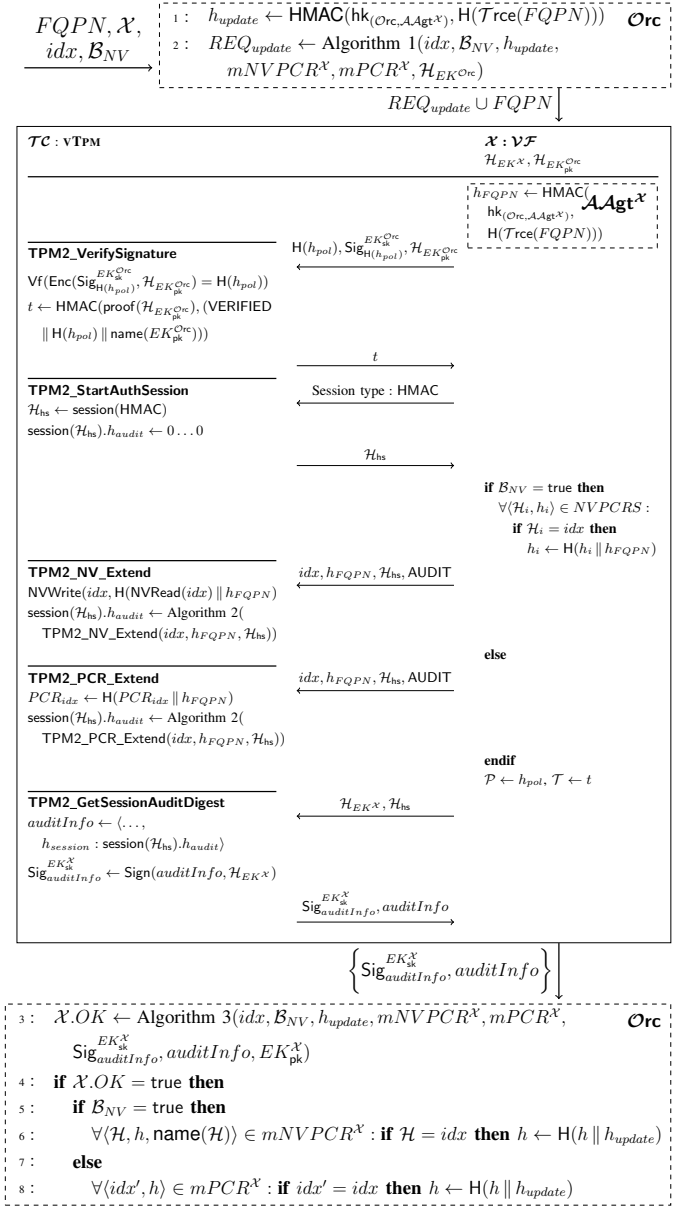

    \centering
    \begin{adjustbox}{
    , width=1\columnwidth}
    \begin{bbrenv}{A}
    
    \begin{bbrbox}[abovesep=0pt]
    \pseudocode[mode=text, colsep=0pt, colspace=0pt, jot=0mm]{
    $\bm{\mathcal{TC}} : \textsc{\textbf{vTpm}}$ \>\> $\bm{\mathcal{X}: \vf}$\\
    \>\> $\mathcal{H}_{EK^\mathcal{X}}, \mathcal{H}_{EK^{\orc}_\pk}$ \\[][\bottomrule] \\[-0.5\baselineskip]
    \>\> \\
    \>\> \\
    \> \sendmessage{<-}{style={line width=\lightrulewidth}, top={$\hash(h_{pol}), \sig^{EK^{\orc}_\sk}_{\hash(h_{pol})}, \mathcal{H}_{EK^{\orc}_\pk}$}} \> \\[-2.2\baselineskip]
    \xdash[17.5em] \>\> \\[-0.5\baselineskip]
    $\bm{\mathsf{TPM2\_VerifySignature}}$ \>\> \\
    $\verify(\enc(\sig^{EK^{\orc}_\sk}_{\hash(h_{pol})}, \mathcal{H}_{EK^{\orc}_\pk}) = \hash(h_{pol}))$ \>\> \\
    $t \gets \mathsf{HMAC}(\mathsf{proof}(\mathcal{H}_{EK^{\orc}_\pk}), (\mathsf{VERIFIED}$ \>\> \\
    $\quad \concat \, \hash(h_{pol}) \, \concat \, \mathsf{name}(EK^{\orc}_\pk)))$ \>\> \\
    \> \sendmessage{->}{style={line width=\lightrulewidth}, top={$t$}} \> \\
    \> \sendmessage{<-}{style={line width=\lightrulewidth}, top={$\text{Session type}: \mathsf{HMAC}$}} \> \\[-2.2\baselineskip]
    \xdash[17.5em] \>\> \\[-0.5\baselineskip]
    $\bm{\mathsf{TPM2\_StartAuthSession}}$ \>\> \\
    $\mathcal{H}_\mathsf{hs} \gets \mathsf{session}(\mathsf{HMAC})$ \>\> \\
    $\mathsf{session}(\mathcal{H}_\mathsf{hs}) . h_{audit} \gets 0\ldots0$ \>\> \\
    \> \sendmessage{->}{style={line width=\lightrulewidth}, top={$\mathcal{H}_\mathsf{hs}$}} \> \\
    \>\> $\textbf{if } \mathcal{B}_{NV} = \true \textbf{ then}$ \\
    \>\> \pcind $\forall \langle \mathcal{H}_i, h_i \rangle \in NVPCRS :$ \\
    \>\> \pcind\pcind $\textbf{if } \mathcal{H}_i = idx \textbf{ then }$ \\
    \>\> \pcind\pcind $\quad h_i \gets \hash(h_i \, \concat \, h_{FQPN})$ \\
    \> \sendmessage{<-}{style={line width=\lightrulewidth}, top={$idx, h_{FQPN}, \mathcal{H}_\mathsf{hs}, \mathsf{AUDIT}$}} \> \\[-2.2\baselineskip]
    \xdash[17.5em] \>\> \\[-0.5\baselineskip]
    $\bm{\mathsf{TPM2\_NV\_Extend}}$ \>\> \\
    $\mathsf{NVWrite}(idx, \hash(\mathsf{NVRead}(idx) \, \concat \, h_{FQPN})$ \>\> \\
    $\mathsf{session}(\mathcal{H}_\mathsf{hs}) . h_{audit} \gets \text{Algorithm~\ref{alg:witness}}($ \>\> \\
    $\quad \mathsf{TPM2\_NV\_Extend}(idx, h_{FQPN}, \mathcal{H}_\mathsf{hs}))$ \>\> \\
    \>\> \textbf{else} \\
    \> \sendmessage{<-}{style={line width=\lightrulewidth}, top={$idx, h_{FQPN}, \mathcal{H}_\mathsf{hs}, \mathsf{AUDIT}$}} \> \\[-2.2\baselineskip]
    \xdash[17.5em] \>\> \\[-0.5\baselineskip]
    $\bm{\mathsf{TPM2\_PCR\_Extend}}$ \>\> \\
    $PCR_{idx} \gets \hash(PCR_{idx} \, \concat \, h_{FQPN})$ \>\> \\
    $\mathsf{session}(\mathcal{H}_\mathsf{hs}) . h_{audit} \gets \text{Algorithm~\ref{alg:witness}}($ \>\> \\
    $\quad \mathsf{TPM2\_PCR\_Extend}(idx, h_{FQPN}, \mathcal{H}_\mathsf{hs}))$ \>\> \\
    \>\> \textbf{endif} \\
    \>\> $\mathcal{P} \gets h_{pol}$, $\mathcal{T} \gets t$ \\
    \> \sendmessage{<-}{style={line width=\lightrulewidth}, top={$\mathcal{H}_{EK^\mathcal{X}} , \mathcal{H}_\mathsf{hs}$}} \> \\[-2.2\baselineskip]
    \xdash[17.5em] \>\> \\[-0.5\baselineskip]
    $\bm{\mathsf{TPM2\_GetSessionAuditDigest}}$ \>\> \\
    $auditInfo \gets \langle \ldots,$ \>\> \\
    $\quad h_{session} : \mathsf{session}(\mathcal{H}_\mathsf{hs}) . h_{audit} \rangle$ \>\> \\
    $\sig^{EK^\mathcal{X}_\sk}_{auditInfo} \gets \sign(auditInfo, \mathcal{H}_{EK^\mathcal{X}})$ \>\> \\
    \> \sendmessage{->}{style={line width=\lightrulewidth}, top={$\sig^{EK^\mathcal{X}_\sk}_{auditInfo}, {auditInfo}$}} \>
    }
    \end{bbrbox}

    \node[draw, dashed, line width=\lightrulewidth, above=1.8cm of A.north east, anchor=east] (orc1) {\pseudocode[mode=text, codesize=\large, linenumbering]{
        $h_{update} \gets \textsf{HMAC}(\hk_{(\orc, \aagent^\mathcal{X})}, \hash(\trace(FQPN)))\quad\quad\quad$\\
        $REQ_{update} \gets \text{Algorithm~\ref{alg:policyCreation}}(idx, \mathcal{B}_{NV}, h_{update},$\pcskipln \\
        $\quad mNVPCR^\mathcal{X}, mPCR^\mathcal{X}, \mathcal{H}_{EK^{\orc}})$
    }};
    \node[anchor=north east, xshift=-0.1cm, yshift=-0.1cm] at (orc1.north east) {\large{$\bm{\orc}$}};
    \draw[->, line width=\heavyrulewidth] ([xshift=-2.5cm]orc1.south east)--([xshift=-2.5cm, yshift=0.1cm]A.north east) node [midway, left] {\large $REQ_{update} \cup FQPN$};
    
    \draw[->, line width=\heavyrulewidth] ([xshift=-3cm, yshift=-.5cm]orc1.west)--([xshift=-0.15cm, yshift=-.5cm]orc1.west) node [midway, above] {\Large \shortstack{$FQPN, \mathcal{X},$\\$idx, \mathcal{B}_{NV}$}};

    \node[draw, dashed, line width=\lightrulewidth, anchor=east] (agent) at([yshift=-2cm,xshift=-0.3cm]A.north east) {\pseudocode[mode=text]{
        $h_{FQPN} \gets \mathsf{HMAC}(\quad\quad\quad\quad\,$\\
        $\quad \hk_{(\orc, \aagent^\mathcal{X})},$\\
        $\quad \hash(\trace(FQPN)))$
    }};
    \node[anchor=north east, xshift=-0.1cm, yshift=-0.1cm] at (agent.north east) {\large{$\bm{\aagent^\mathcal{X}}$}};

    \node[draw, dashed, line width=\lightrulewidth, below=3.6cm of A.south east, anchor=east] (orc2) {\pseudocode[mode=text, codesize=\large, linenumbering, lnstart=2]{
        $\mathcal{X}.OK \gets \text{Algorithm~\ref{alg:verifyAuditDigest}}(idx, \mathcal{B}_{NV}, h_{update}, mNVPCR^\mathcal{X}, mPCR^\mathcal{X},\quad\quad$\pcskipln\\
        \pcind $\sig^{EK^\mathcal{X}_\sk}_{auditInfo}, auditInfo, EK^\mathcal{X}_\pk)$\\
        $\textbf{if } \mathcal{X}.OK = \true \textbf{ then }$\\
        \pcind $\textbf{if } \mathcal{B}_{NV} = \true \textbf{ then }$\\
        \pcind \pcind $\forall \langle \mathcal{H}, h, \textsf{name}(\mathcal{H}) \rangle \in mNVPCR^\mathcal{X}: \textbf{if } \mathcal{H} = idx \textbf{ then } h \gets \hash(h \, \concat \, h_{update})$\\
        \pcind $\textbf{else}$\\
        \pcind\pcind $\forall \langle idx^\prime, h \rangle \in mPCR^\mathcal{X}: \textbf{if } idx^\prime = idx \textbf{ then } h \gets \hash(h \, \concat \, h_{update})$
    }};
    \node[anchor=north east, xshift=-0.1cm, yshift=-0.1cm] at (orc2.north east) {\large{$\bm{\orc}$}};
    \draw[->, line width=\heavyrulewidth] ([xshift=-2.5cm]A.south east)--([xshift=-2.5cm, yshift=0.1cm]orc2.north east) node [midway, left] {\large $\set{\sig^{EK^\mathcal{X}_\sk}_{auditInfo}, auditInfo}$};
    
    \end{bbrenv}
    \end{adjustbox}
    \caption{Measurement update}
    \label{fig:update}
\end{figure}

To enforce a configuration update, $\orc$ uses the mock PCRs associated with $\mathcal{X}$ to emulate what the \emph{expected} (thereby \emph{trusted}) cascading effect of the update's measurement is and includes the result in a new policy. For example, let $r$ be a resource on $\mathcal{X}$ (also known to $\orc$) and $i$ be a PCR attached to $\mathcal{X}$ which will house $r$'s measurement. On $\orc$, the current (mock) value of $i$ is assumed to be $v$. Thus, the expected value in PCR $i$ after measuring $r$ is $\hash(v \, \concat \, \hash(r))$. 
The measurement-update protocol is shown in Fig.~\ref{fig:update}, where, given a Fully Qualified Path Name (FQPN) of some configuration on a $\mathcal{VF}$ ($\mathcal{X}$) and a target PCR ($idx$), $\orc$ locally measures and authenticates (using the shared secret between $\orc$ and $\mathcal{X}$'s $\aagent$) the configuration measurement and then applies Algorithm~\ref{alg:policyCreation} to compose and authorize the expected policy digest using the mock PCRs associated with $\mathcal{X}$ (described in Section~\ref{subsubsec:securityAnalysisFreshness}). The authorized policy and details for $\mathcal{X}$ to perform the measurement locally (i.e., FQPN, PCR type, and $idx$) are then sent to $\mathcal{X}$. On $\mathcal{X}$, $\aagent^\mathcal{X}$ intercepts the update request, measures FQPN using $\trace$, and authenticates the measurement. $\mathcal{X}$ then proceeds to use its vTPM to verify whether the supplied policy digest was signed using $\orc$'s EK. If the signature is correct, the vTPM returns a ticket denoting that the vTPM has verified the policy digest's correctness. 

To prove to $\orc$ that the correct PCR ($idx$) was extended, $\mathcal{X}$ starts an HMAC session and runs the extend command in audit mode to have the vTPM internally witness (see Algorithm~\ref{alg:witness}) the incoming CPs and outgoing Response Parameters (RP) into the session's audit digest ($cpHash,\allowbreak rpHash,\allowbreak auditDigest$ are described in Part 1 of the TPM 2.0 specifications~\cite{tcgTpmArchitecture}). Once the command completes, $\mathcal{X}$ asks the vTPM to certify the current session's audit digest with $\mathcal{X}$'s EK and sends it to $\orc$. To verify the audit digest (Algorithm~\ref{alg:verifyAuditDigest}), $\orc$ first computes the expected audit digest, with the correct arguments and a successful Response Code (RC). If $\mathcal{X}$'s audit digest differs from the expected, or the signature is incorrect, $\mathcal{X}$ did poorly. 

\normalsize

\begin{algorithm}[htbp]
    \DontPrintSemicolon
    \SetKwInOut{Input}{Input}
    \SetKwInOut{Output}{Output}
    \Input{$idx, \mathcal{B}_{NV}, h_{update}, mNVPCR, mPCR, \mathcal{H}_{EK}$}
    \Output{$\set{h_{pol}, \hash(h_{pol}), \sig^{\key}_{\hash(h_{pol})}, idx, \mathcal{B}_{NV}}$}
    \uIf{$\mathcal{B}_{NV} = \true$}
    {
        $\forall \langle \mathcal{H}, h, \mathsf{name}(\mathcal{H}) \rangle \in mNVPCR:$\;
        \Indp$\textbf{if } \mathcal{H} = idx \textbf{ then } h \gets \hash(h \, \concat \, h_{update})$\;
    }
    \Else{
        $\forall \langle idx^\prime, h \rangle \in mPCR:$\;
        \Indp$\textbf{if } idx^\prime = idx \textbf{ then } h \gets \hash(h \, \concat \, h_{update})$\;
    }
    $h_\text{pol} \gets 0\ldots0$\;
    $\forall \langle \mathcal{H}, h, \mathsf{name}(\mathcal{H}) \rangle \in mNVPCR:$\;
    \Indp$args \gets \hash(h \, \concat \, 0x0000 \, \concat \, 0x0000)$\;
    $h_{pol} \gets \hash(h_{pol} \, \concat \, CC_{PolicyNV} \, \concat \, args \, \concat \, \mathsf{name}(\mathcal{H}))$\;
    \Indm\If{$mPCR \neq \emptyset$}
    {
        $h_{PCR} \gets \emptyset, indices \gets \emptyset$\;
        $\forall \langle idx^\prime, h \rangle \in mPCR:$\;
        \Indp$h_{PCR} \gets h_{PCR} \, \concat \, h$\;
        $indices \gets indices \cup idx^\prime$\;
        \Indm$h_{pol} \gets \hash(h_{pol} \concat CC_{PolicyPCR} \concat indices \concat \hash(h_{PCR}))$
    }
    $\sig^{EK_{\sk}}_{\hash(h_{pol})} \gets tpm.\sign(\hash(h_{pol}), \mathcal{H}_{EK})$\;
    \textbf{return} $h_{pol}, \hash(h_{pol}), \sig^{EK_{\sk}}_{\hash(h_{pol})}, idx, \mathcal{B}_{NV}$\;
    \caption{Composing AK policy update requests}
    \label{alg:policyCreation}
\end{algorithm}

\begin{algorithm}[htbp]
    \DontPrintSemicolon
    \SetKwInOut{Input}{Input}
    \SetKwInOut{Output}{Output}
    \Input{$\mathsf{CMD} : \mathcal{H}_0 \times \mathcal{H}_1 \times \mathcal{H}_2 \times params \rightarrow RC \times CC \times rparams$}
    \Output{$h_{audit}^\prime$ - updated audit session digest}
    $cpHash \gets \hash(CC(\mathsf{CMD}) \, \concat \, \mathsf{name}(\mathcal{H}_0) \, \concat \, \mathsf{name}(\mathcal{H}_1)$ $\concat \, \mathsf{name}(\mathcal{H}_2) \, \concat \, params)$\;
    $rpHash \gets \hash(RC(\eval(\mathsf{CMD})) \, \concat \, CC_{\mathsf{CMD}} \, \concat \, rparams)$\;
    $h_{audit}^\prime \gets \hash(h_{audit} \, \concat \, cpHash \, \concat \, rpHash)$\;
    \textbf{return} $h_{audit}^\prime$
    \caption{Witness}
    \label{alg:witness}
\end{algorithm}

\begin{algorithm}[htbp]
    \DontPrintSemicolon
    \SetKwInOut{Input}{Input}
    \SetKwInOut{Output}{Output}
    \Input{$idx, \mathcal{B}_{NV}, h_{update}, mNVPCR, mPCR,$ $\sig^{EK_\sk}_{auditInfo}, auditInfo, EK_\pk$}
    \Output{$\mathcal{B}$}
    \uIf{$\mathcal{B}_{NV} = \true$}
    {
        $\forall \langle \mathcal{H}, h, \mathsf{name}(\mathcal{H}) \rangle \in mNVPCR:$\;
        \Indp $\textbf{if } \mathcal{H} = idx \textbf{ then}$\;
        \Indp $cpHash \gets \hash(CC_{NV\_Extend} \, \concat \, \mathsf{name}(\mathcal{H})$ $\concat \, \mathsf{name}(\mathcal{H}) \, \concat \, \mathsf{len}(h_{update}) \, \concat \, h_{update})$\;
        $rpHash \gets \hash(\textsf{success} \, \concat \, CC_{NV\_Extend})$\;
    }
    \Else{
        $\forall \langle idx^\prime, h \rangle \in mPCR:$\;
        \Indp$\textbf{if } idx^\prime = idx \textbf{ then}$\;
        \Indp$cpHash \gets \hash(CC_{PCR\_Extend} \, \concat \, idx^\prime \, \concat \, idx^\prime$ $\concat \, authHash \, \concat \, h_{update})$\;
        $rpHash \gets \hash(\textsf{success} \, \concat \, CC_{PCR\_Extend})$\;
    }
    $h_{audit} \gets \hash(0\ldots0 \, \concat \, cpHash \, \concat \, rpHash)$\;
    $\verify(h_{audit} = auditInfo . h_{session})$\;
    $\verify(\enc(\sig^{EK_\sk}_{auditInfo}, EK_\pk) = auditInfo)$\;
    \textbf{return} $\true$\;
    \caption{Verify session audit digest}
    \label{alg:verifyAuditDigest}
\end{algorithm}

\subsubsection{Proof of Conformance}
\label{subsubsec:proofOfConformance}

\begin{figure}[htbp]
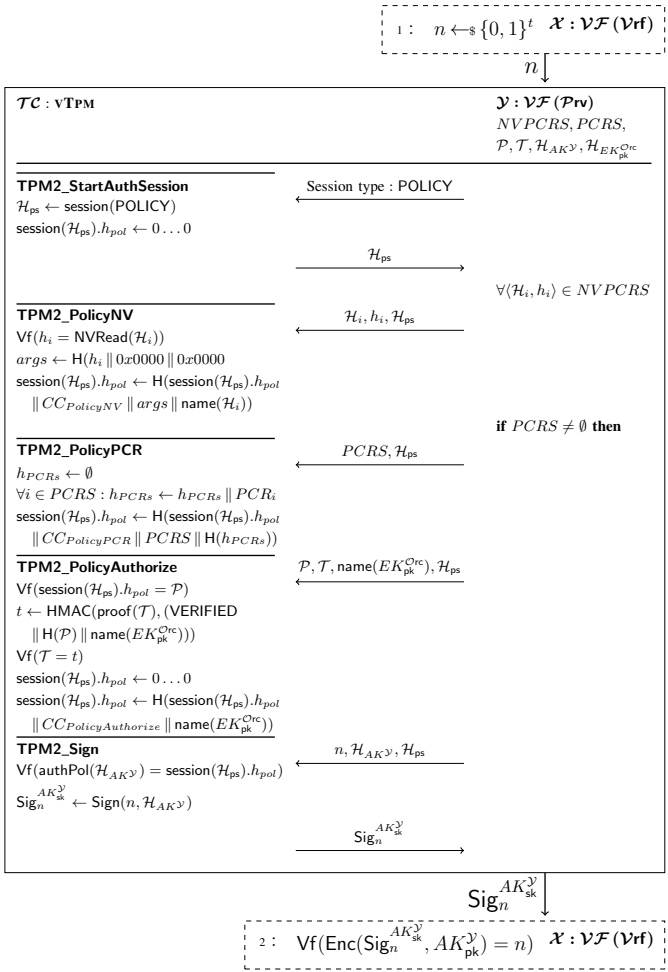

    \centering
    \begin{adjustbox}{
    , width=1\columnwidth}
    \begin{bbrenv}{B}
    
    \begin{bbrbox}[abovesep=0pt]
    \pseudocode[mode=text, colsep=0cm, colspace=0cm, jot=0mm]{
    $\bm{\mathcal{TC}} : \textsc{\textbf{vTpm}}$ \>\> $\bm{\mathcal{Y}: \vf\, (\prover)}$ \\
    \>\> $NVPCRS, PCRS,$ \\
    \>\> $\mathcal{P}, \mathcal{T}, \mathcal{H}_{AK^\mathcal{Y}}, \mathcal{H}_{EK^{\orc}_\pk}$\\[][\bottomrule] \\[-0.5\baselineskip]
    \> \sendmessage{<-}{style={line width=\lightrulewidth}, top=$\text{Session type}: \mathsf{POLICY}$} \> \\[-2.2\baselineskip]
    \xdash[17em] \>\> \\[-0.5\baselineskip]
    $\bm{\mathsf{TPM2\_StartAuthSession}}$ \>\> \\
    $\mathcal{H}_\mathsf{ps} \gets \mathsf{session}(\mathsf{POLICY})$ \>\> \\
    $\mathsf{session}(\mathcal{H}_\mathsf{ps}) . h_{pol} \gets 0\ldots0$ \>\> \\
    \> \sendmessage{->}{style={line width=\lightrulewidth}, top=$\mathcal{H}_\mathsf{ps}$} \> \\
    \>\> $\forall \langle \mathcal{H}_i, h_i \rangle \in NVPCRS$ \\
    \> \sendmessage{<-}{style={line width=\lightrulewidth}, top={$\mathcal{H}_i, h_i, \mathcal{H}_\mathsf{ps}$}} \> \\[-2.2\baselineskip]
    \xdash[17em] \>\> \\[-0.5\baselineskip]
    $\bm{\mathsf{TPM2\_PolicyNV}}$ \>\> \\
    $\verify(h_i = \mathsf{NVRead}(\mathcal{H}_i))$ \>\> \\
    $args \gets \hash(h_i \, \concat \, 0x0000 \, \concat \, 0x0000$ \>\> \\
    $\mathsf{session}(\mathcal{H}_\mathsf{ps}) . h_{pol} \gets \hash(\mathsf{session}(\mathcal{H}_\mathsf{ps}) . h_{pol}$ \>\> \\
    $\quad \concat \, CC_{PolicyNV} \, \concat \, args \, \concat \, \mathsf{name}(\mathcal{H}_i))$ \>\> \\
    \>\> $\textbf{if } PCRS \neq \emptyset \textbf{ then}$ \\
    \> \sendmessage{<-}{style={line width=\lightrulewidth}, top={$PCRS, \mathcal{H}_\mathsf{ps}$}} \> \\[-2.2\baselineskip]
    \xdash[17em] \>\> \\[-0.5\baselineskip]
    $\bm{\mathsf{TPM2\_PolicyPCR}}$ \>\> \\
    $h_{PCRs} \gets \emptyset$ \>\> \\
    $\forall i \in PCRS : h_{PCRs} \gets h_{PCRs} \, \concat \, PCR_i$ \>\> \\
    $\mathsf{session}(\mathcal{H}_\mathsf{ps}) . h_{pol} \gets \hash(\mathsf{session}(\mathcal{H}_\mathsf{ps}) . h_{pol}$ \>\> \\
    $\quad \concat \, CC_{PolicyPCR} \, \concat \, PCRS \, \concat \, \hash(h_{PCRs}))$ \>\> \\
    \> \sendmessage{<-}{style={line width=\lightrulewidth}, top={$\mathcal{P}, \mathcal{T}, \mathsf{name}(EK^{\orc}_\pk), \mathcal{H}_\mathsf{ps}$}} \> \\[-2.2\baselineskip]
    \xdash[17em] \>\> \\[-0.5\baselineskip]
    $\bm{\mathsf{TPM2\_PolicyAuthorize}}$ \>\> \\
    $\verify(\mathsf{session}(\mathcal{H}_\mathsf{ps}) . h_{pol} = \mathcal{P})$ \>\> \\
    $t \gets \mathsf{HMAC}(\mathsf{proof}(\mathcal{T}), (\mathsf{VERIFIED}$ \>\> \\
    $\quad \concat \, \hash(\mathcal{P}) \, \concat \, \mathsf{name}(EK^{\orc}_\pk)))$ \>\> \\
    $\verify(\mathcal{T} = t)$ \>\> \\
    $\mathsf{session}(\mathcal{H}_\mathsf{ps}) . h_{pol} \gets 0\ldots0$ \>\> \\
    $\mathsf{session}(\mathcal{H}_\mathsf{ps}) . h_{pol} \gets \hash(\mathsf{session}(\mathcal{H}_\mathsf{ps}) . h_{pol}$ \>\> \\
    $\quad \concat \, CC_{PolicyAuthorize} \, \concat \, \mathsf{name}(EK^{\orc}_\pk))$ \>\> \\
    \> \sendmessage{<-}{style={line width=\lightrulewidth}, top={$n, \mathcal{H}_{AK^\mathcal{Y}}, \mathcal{H}_\mathsf{ps}$}} \> \\[-2.2\baselineskip]
    \xdash[17em] \>\> \\[-0.5\baselineskip]
    $\bm{\mathsf{TPM2\_Sign}}$ \>\> \\
    $\verify(\mathsf{authPol}(\mathcal{H}_{AK^\mathcal{Y}}) = \mathsf{session}(\mathcal{H}_\mathsf{ps}) . h_{pol})$ \>\> \\
    $\sig^{AK^\mathcal{Y}_\sk}_n \gets \sign(n, \mathcal{H}_{AK^\mathcal{Y}})$ \>\> \\
    \> \sendmessage{->}{style={line width=\lightrulewidth}, top={$\sig^{AK^\mathcal{Y}_\sk}_n$}} \>
    }
    \end{bbrbox}

    \node[draw, dashed, line width=\lightrulewidth, above=1.2cm of B.north east, anchor=east, minimum height=1cm] (vf1) {\pseudocode[mode=text, codesize=\large, linenumbering]{
        $n \sample \bin^t\quad\quad\quad\quad\quad\quad$
    }};
    \node[anchor=north east, xshift=-0.1cm, yshift=-0.1cm] at (vf1.north east) {\normalsize{$\bm{\mathcal{X}: \vf\, (\verifier)}$}};
    \draw[->, line width=\heavyrulewidth] ([xshift=-2.5cm]vf1.south east)--([xshift=-2.5cm, yshift=0.1cm]B.north east) node [midway, left] {\Large $n$};;
    
    \node[draw, dashed, line width=\lightrulewidth, below=1.5cm of B.south east, anchor=east, minimum height=1cm] (vf2) {\pseudocode[mode=text, codesize=\large, linenumbering, lnstart=1]{
        $\verify(\enc(\sig^{AK^\mathcal{Y}_\sk}_n, AK^\mathcal{Y}_\pk) = n)\quad\quad\quad\quad\quad\quad$
    }};
    \node[anchor=north east, xshift=-0.1cm, yshift=-0.1cm] at (vf2.north east) {\normalsize{$\bm{\mathcal{X}: \vf\, (\verifier)}$}};
    \draw[->, line width=\heavyrulewidth] ([xshift=-2.5cm]B.south east)--([xshift=-2.5cm, yshift=0.1cm]vf2.north east) node [midway, left] {\Large $\sig^{AK^\mathcal{Y}_\sk}_n$};
    
    \end{bbrenv}
    \end{adjustbox}
    \caption{Oblivious Remote Attestation (ORA)}
    \label{fig:ora}
\end{figure}

Equipped with an authorized policy, $\mathcal{X}$ can serve attestation requests. When another $\vf$, $\mathcal{Y}$, wants to determine whether $\mathcal{X}$ is correct, $\mathcal{Y}$ sends $\mathcal{X}$ a nonce $n$. If $\mathcal{X}$ responds with a signature over $n$ using its certified AK, then $\mathcal{Y}$ knows that $\mathcal{X}$ fulfills $\orc$'s requirements. The sequence of steps performed by $\mathcal{X}$ are shown in Fig.~\ref{fig:ora}, where $\mathcal{X}$ first executes a series of policy commands (i.e., \texttt{PolicyPCR} and \texttt{PolicyNV}) to verify and measure the currently active PCRs (Section~\ref{subsubsec:remotePcrAdministration}) in a session's policy digest. Once all PCRs have been accounted for, $\mathcal{X}$ runs \texttt{PolicyAuthorize} with the verified ticket (Section~\ref{subsubsec:supervisedUpdates}) and authorized policy (denoted $\mathcal{P}$). If the session's policy digest corresponds to the approved policy, then the vTPM replaces the session's policy digest with the name (digest over the public area) of $\orc$'s EK, which allows $\mathcal{X}$ to wield its AK and sign $\mathcal{Y}$'s challenge.

\normalsize

\subsection{Implementation}

We implemented the protocols in C++ and tested them on two platforms: one with a SW-TPM and another with a HW-TPM (see Appendix~\ref{subsec:environmentalSetup}). The protocols were benchmarked on both platforms, and the results are presented in Appendix~\ref{appendix:timings}.

\section{Security Analysis}

\subsection{Security Properties}

We proceed to evaluate how our scheme upholds the security properties (Section~\ref{subsec:highLevelSecurityProperties}) under the considered threat model.

\subsubsection{Property~\ref{property:configurationCorrectness}: Configuration Correctness}

Let $\adv$ extend the PCRs (see Fig.~\ref{fig:update}) with measurements of her choice during a measurement update, and $\alpha$ be the configuration that $\orc$ has requested to be measured. If $\alpha$ was altered without first recording its digest, $d \gets \hash(\alpha)$, $\adv$ cannot win unless she picks a random digest $d^\prime$, where $d^\prime = d$. If $\alpha$ is unchanged, $\adv$ computes $d \gets \hash(\alpha)$ and supplies $d$. However, since $\alpha$ is correct, Property~\ref{property:configurationCorrectness} is not violated. Now, assume that $\adv$ altered $\alpha$, but her chosen $d^\prime$ is correct. Her next challenge is to guess $\aagent$'s secret ($\hk$) to solve for $\mathsf{HMAC}(\hk, d^\prime)$. Unless she solves this challenge, she cannot extend the correct measurement, and verification of the session's audit digest will fail on $\orc$. Thus, circumventing the measurement process's integrity is infeasible, and thus the property holds. Further, since metadata is included in measurements (Section~\ref{subsec:adversarialModel}), $\adv$ cannot unnoticeably alter and restore configurations between updates. However, although not covered by the property, alterations to the configurations currently remain undetected until the next measurement. We propose two directions to mitigate this Time-Of-Check to Time-Of-Use (TOCTOU) problem~\cite{nunes2020toctou}.

\paragraph{Reactive (lazy) TOCTOU-resistance}

The first approach is to require $\aagent$ to vouch for the configuration's correctness at the \emph{time of processing the attestation request} by either: (\emph{i}) comparing the metadata (e.g., the \texttt{i\_generation} and \texttt{i\_version} fields), or (\emph{ii}) re-measuring the configuration. However, for this to be useful, considering that $\adv$ can block access to $\aagent$ (Section~\ref{subsec:adversarialModel}), we must extend the existing attestation policies (Section~\ref{subsubsec:supervisedUpdates}) to require proof that $\aagent$ handled the attestation. We can achieve this with \texttt{PolicyAuthValue} and require that $\hk_{(\orc, \aagent)}$ be supplied (along with the necessary \texttt{PolicyNV} and \texttt{PolicyPCR} commands) for 
\texttt{PolicyAuthorize} to succeed (see Fig.~\ref{fig:ora}). Note, however, since \texttt{PolicyAuthValue} does not support limiting when authorization should expire, $\aagent$ must close the policy session's handle once it has signed the $\verifier$'s challenge. If $\aagent$ had an asymmetric key pair, we could have instead used \texttt{PolicySigned}, which allows specifying when authorization to the AK expires, like a ``dead man's switch''. 

\paragraph{Proactive TOCTOU-resistance}

Another approach is to extend $\aagent$ software to continuously monitor all objects in $FQPN \in REQ_{update}$ between updates. If the configuration changes, $\aagent$ effectively neuters the $\vf$'s AK by extending the \emph{active} PCRs. However, to achieve this efficiently is non-trivial. The most notable framework is the conjunction of IMA (Section~\ref{sec:relatedWork}) and Extended Verification Module (EVM), which for the Linux-based kernels, provide fine-grained mechanisms to measure and detect file alterations. However, since IMA lacks support to change MP during run-time, it is unfit in our case. Another increasingly popular method, also in the context of containerization security~\cite{findlay2021bpfcontain}, is the use of extended Berkeley Packet Filters (eBPF). With eBPF, extensions can be applied to the OS kernel during run-time, enabling (privileged) software to hook and filter system calls dynamically. Employing the bpftrace~\cite{bpftrace} tool or BPF Compiler Collection (BCC) toolkit, we can instrument $\aagent$ to attach hooks (or \emph{probes}) on file-related system calls and match calls targeting the configurations. For example, to detect writes and deletions we can attach \texttt{sys\_enter\_write} and \texttt{vfs\_unlink} probes, and to catch calls that open configuration files in modes other than \emph{read-only}, we can leverage \texttt{sys\_enter\_openat}. Note, however, that additional probes are required in practice since files can also be written in other ways (e.g., using \texttt{mmap}). Nonetheless, utilizing eBPF, we can effectively and preemptively mitigate the TOCTOU problem.

\subsubsection{Secure Enrollment (Property~\ref{property:remoteDeployment})}
\label{subsubsec:secureEnrollment}

To ensure that $\orc$ controls the use of all AKs, they must be created to only abide by policies signed by $\orc$. Since an AK must be certified by the $\vf$'s EK (using \texttt{CertifyCreation}) to be accepted by $\orc$, where EK, for all $\vf$s, is a credentialed non-duplicable EK (\emph{restricted} signing key) that can only sign TPM-generated data, $\adv$ can neither fool $\orc$ to accept a self-signed creation certificate nor have the EK sign a $\adv$-forged certificate. Also, if any details in the AK's certificate (e.g., its attributes, name, or authorization policy) are incorrect, $\orc$ rejects it. Thus, $\adv$ cannot threaten the AK creation process's integrity. Note that \emph{forward acceptance} (Property~\ref{property:forwardAcceptance}) is ensured during  AK creation by requiring that the authorization policy be a flexible policy bound to $\orc$'s public EK. Thus, Property~\ref{property:remoteDeployment} $\implies$ Property~\ref{property:forwardAcceptance}.

\subsubsection{Freshness (Property~\ref{property:policyFreshness})}
\label{subsubsec:securityAnalysisFreshness}

Given a configuration update $h_{update}$ for a $\vf$, $\mathcal{X}$, $\orc$ uses Algorithm~\ref{alg:policyCreation} (Fig.~\ref{fig:update}) with $\mathcal{X}$'s \emph{current} mock structures, $mPCR, mNVPCR$, to compose a policy which accounts for the update. The algorithm performs the following actions: (\emph{i}) accumulate $h_{update}$ into the appropriate mock PCR (lines 1 to 7); (\emph{ii}) initialize a policy $h_{pol}$ (line 8); (\emph{iii}) extend $h_{pol}$ with a simulation, where, for each mocked NV PCR $i$, \texttt{PolicyNV} is executed with $i$'s current measurement (lines 8 to 11); (\emph{iv}) if nonempty, extend $h_{pol}$ with a simulation, where all PCRs in $mPCR$ are selected and their accumulated digest is supplied to \texttt{PolicyPCR} (lines 12 to 18); (\emph{v}) sign $\hash(h_{pol})$. The signature and $h_{pol}$ are then sent to $\mathcal{X}$, where $\adv$ also has access to it. Given the authorized $h_{pol}$, AK is unlocked using \texttt{PolicyAuthorize} if, after executing the \emph{exact same} sequence of commands, the vTPM's internally accumulated digest $h_{pol}^\prime$ is equal to the authorized digest: $h_{pol}^\prime = h_{pol}$.

When, at a later time, $\mathcal{X}$ must account for another update, $h_{update}^\prime$, its \emph{current} mock structures $mPCR^\prime, mNVPCR^\prime$ are again used to authorize a new policy digest $h_{pol}^{\prime\prime}$. However, if $\{indicies(mPCR^\prime) \cap indicies(mPCR)\} \cup \{indicies(\allowbreak mNVPCR)^\prime \cap indicies(mNVPCR)\} = \emptyset$, then $h_{pol}$ and $h_{pol}^{\prime\prime}$ share no elements (PCRs), which means that both policies can simultaneously unlock $\mathcal{X}$'s AK. Thus, when another $\vf$, $\mathcal{Y}$, wants to verify $\mathcal{X}$'s correctness, it is undefined \emph{which} policy it fulfills. It is therefore necessary that policies are either (\emph{i}) created with \emph{at least one} element in common with the preceding policy and that this element be extended to neuter the preceding policy, or (\emph{ii}) followed by another command which extends one PCR of the preceding policy.

\subsubsection{Property~\ref{prop:zeroKnowledgeVerification}: Zero-Knowledge CIV}

When a $\vf$, $\mathcal{X}$, who knows $\orc$'s identity and public EK, first learns about another $\vf$, $\mathcal{Y}$, it receives $\{\sig^{EK^{\orc}_{\sk}}_{AK^{\mathcal{Y}}_{\pk}}, AK^{\mathcal{Y}}_{\pk}\}$ from $\orc$ (Fig.~\ref{fig:workFlow}). If $\text{Property~\ref{property:configurationCorrectness}} \land \text{Property~\ref{property:remoteDeployment}} \land \text{Property~\ref{property:policyFreshness}}$ hold, then $\mathcal{Y}$ is correctly associated with $AK^{\mathcal{Y}}_{\pk}$. Thus, if $\mathcal{X}$ chooses a random number $n \sample \bin^t$ and $\mathcal{Y}$ presents $\sig^{AK^{\mathcal{Y}}_{\sk}}_n$, where $\enc\large(\sig^{AK^\mathcal{Y}_\sk}_n, AK^\mathcal{Y}_\pk) = n$, then $\mathcal{X}$ knows that $\mathcal{Y}$ was able to use $AK^{\mathcal{Y}}_{\sk}$ and thence fulfills $\orc$'s requirements. Thus, $\text{Property~\ref{property:configurationCorrectness}} \land \text{Property~\ref{property:remoteDeployment}} \land \text{Property~\ref{property:policyFreshness}} \implies \text{Property~\ref{prop:zeroKnowledgeVerification}}$.

\section{Conclusions}
\label{sec:conclusions}

In this work, we presented an architecture to support confidential CIV using well-known, trusted computing techniques. With this solution, trust-aware $\mathcal{SG}$ chains can be created with verifiable evidence on the integrity assurance and correctness of the comprised containers: from the trusted launch (enrollment) and configuration to the run-time attestation of low-level configuration properties. The proposed scheme considered state-of-the-art remote attestation variants and addressed one of the main challenges concerning assumptions on the $\verifier$ entity's trustworthiness, thus, enabling privacy-preserving integrity correctness. 

\vspace{-0.2cm}
\section{Acknowledgment}
\label{sec:ack}
\vspace{-0.2cm}
This work was supported by the European Commission, under the ASTRID project; Grant Agreements no. 786922.
\vspace{-0.5cm}










\bibliographystyle{IEEEtran}
\bibliography{IEEEabrv,bibs}

\clearpage
\appendix
\subsection{Protocols for Attaching and Detaching PCRs}
\label{appendix:protocols}

Fig.~\ref{fig:createNvPcr} contains the message exchanges to secure the process of attaching PCRs to a $\vf$ ($\mathcal{X}$). For normal PCRs, $\mathcal{X}$ is informed about which PCR to use (track) and both $\orc$ and $\mathcal{X}$ add to their knowledge, i.e., $\orc$ adds it to $mPCR^\mathcal{X}$  and $\mathcal{X}$ to $PCRS$. Otherwise, for NV-based PCRs, $\orc$ sends: (\emph{i}) a NV identifier, (\emph{ii}) a NV template which describes the $\hash$ algorithm and attributes of the NV slot, e.g., that modifications must happen using \texttt{TPM2\_NV\_Extend} (to imitate a PCR), and that a policy is required to delete the index, (\emph{iii}) an authorization policy which gives $\orc$ the exclusive right to authorize the deletion of the index (described Section~\ref{subsubsec:remotePcrAdministration}), and (\emph{iv}) an initial value (IV) to extend. The IV is necessary since newly-created NV indices cannot be read (or certified) until they have been written. Thus, to allow $\mathcal{X}$ to certify it, $\orc$ sends an initial (deterministic) IV which $\mathcal{X}$ must extend the newly created NV-based PCR (NVPCR) with. Once the NVPCR is created, extended, and certified, $\mathcal{X}$ sends the certification details to $\orc$, who verifies that: (\emph{i}) the certified information is of a TPM generated structure, (\emph{ii}) the NVPCR's value is $\hash(0\ldots0 \, \concat \, IV)$, (\emph{iii}) the NVPCR's name is as expected (i.e., that it contains the specified attributes and is bound to the correct authorization policy), (\emph{iv}) the certificate is authentic. If everything holds, then the NVPCR is added to $mNVPCR^\mathcal{X}$ on $\orc$.

To detach a normal PCR, $\orc$ simply informs $\mathcal{X}$ about which PCR to remove from its $PCRS$. For a NVPCR, however, the process is more tricky. To detach a NVPCR, $\orc$ requests $\mathcal{X}$ to start a fresh policy session and return the session's TPM-generated nonce ($n$). With $n$ and one of $\mathcal{X}$'s NVPCRs ($idx$), $\orc$ runs Algorithm~\ref{alg:authorizedNvDeletion}, which: (\emph{i}) authorizes a policy ($h_{pol}$) requiring that \texttt{TPM2\_PolicySigned} be executed with a digest ($aHash$) signed by $\orc$ (lines 1 and 2), (\emph{ii}) signs $aHash$ (as described in Part 2 of the TPM 2.0 specifications~\cite{tcgTpmArchitecture}), which is over $n$, an expiration (set to 0), and a CP digest, $h_{cp}$ ($cpHash$ in Algorithm~\ref{alg:witness}), where $h_{cp}$ restricts the session to the undefine command (as required by the NV index's authorization policy, see Figure~\ref{fig:createNvPcr}) with $idx$ as a parameter (lines 3 to 8). Thus, to perform the deletion, $\mathcal{X}$: (\emph{i}) verifies that $h_{pol}$ was signed by $\orc$, (\emph{ii}) executes \texttt{TPM2\_PolicySigned} which: (\emph{ii-a}) updates the session's policy digest to indicate that the command was executed with some digest signed by $\orc$, and (\emph{ii-b}) sets the session's $cpHash$ to $h_{cp}$, (\emph{iii}) runs \texttt{TPM2\_PolicyAuthorize} with $h_{pol}$, which, if it matches the session's policy digest, sets the session's digest to state that a policy authorized by $\orc$ was fulfilled, (\emph{iv}) runs \texttt{TPM2\_PolicyCommandCode} to restrict the session's CC, (\emph{v}) runs \texttt{TPM2\_NV\_UndefineSpaceSpecial} which deletes the NV index if everything holds, (\emph{vi}) removes the NV index from its local knowledge.

Note that the nonce ($n$) is just a random and unauthenticated number, and the authorized policy generated by $\orc$ carries no information that restricts it to $\mathcal{X}$'s vTPM. Thus, if two vTPMs $A,B$ have the same NV index defined (with the same attributes and bound to the same authorization policy), then the session could belong to either $A$ or $B$, and the authorized policy would succeed. There are two easy solutions to this issue: (\emph{i}) create an \emph{authentic} channel between $\mathcal{X}$ and $\orc$ using software, or (\emph{ii}) include additional (unique) data when creating a NV index such that an authorized policy is unique to a specific $\vf$.

\begin{figure}[htbp]
    \centering
    \begin{adjustbox}{
    , width=1\columnwidth}
    \begin{bbrenv}{D}
    
    \begin{bbrbox}[abovesep=0pt]
    \pseudocode[mode=text, colsep=0pt, colspace=0pt, jot=0mm]{
    $\bm{\mathcal{TC}} : \textsc{\textbf{vTpm}}$ \>\> $\bm{\mathcal{X}: \vf}$\\
    $PPS$ \>\> $\mathcal{H}_{EK^\mathcal{X}}, PCRS, NVPCRS$ \\ [][\bottomrule] \\[-0.5\baselineskip]
    \>\> $\textbf{if } \mathcal{B}_{NV} = \false \textbf{ then}$ \\
    \>\> \pcind $PCRS \gets PCRS \cup idx$ \\
    \>\> $\textbf{else }$ \\
    \> \sendmessage{<-}{style={line width=\lightrulewidth}, top={$\mathcal{H}_{PPS}, idx, \mathsf{TPL}(idx), h_{pol}$}} \> \\[-2.2\baselineskip]
    \xdash[17.5em] \>\> \\[-0.5\baselineskip]
    $\bm{\mathsf{TPM2\_NV\_DefineSpace}}$ \>\> \\
    $\mathsf{NVCreate}(idx, \mathcal{H}_{PPS}, \mathsf{TPL}(idx), h_{pol})$ \>\> \\
    \> \sendmessage{<-}{style={line width=\lightrulewidth}, top={$idx, IV$}} \> \\[-2.2\baselineskip]
    \xdash[17.5em] \>\> \\[-0.5\baselineskip]
    $\bm{\mathsf{TPM2\_NV\_Extend}}$ \>\> \\
    $\mathsf{NVWrite}(idx) \gets \hash(\mathsf{NVRead}(idx) \, \concat \, IV)$ \>\> \\
    \> \sendmessage{<-}{style={line width=\lightrulewidth}, top={$\mathcal{H}_{PPS}, idx, \mathcal{H}_{EK^\mathcal{X}}$}} \> \\[-2.2\baselineskip]
    \xdash[17.5em] \>\> \\[-0.5\baselineskip]
    $\bm{\mathsf{TPM2\_NV\_Certify}}$ \>\> \\
    $certInfo \gets \langle \ldots \rangle$ \>\> \\
    $\sig^{EK^\mathcal{X}_\sk}_{certInfo} \gets \sign(certInfo, \mathcal{H}_{EK^\mathcal{X}})$ \>\> \\[-.75\baselineskip]
    \> \sendmessage{->}{style={line width=\lightrulewidth}, top={$\sig^{EK^\mathcal{X}_\sk}_{certInfo}, certInfo$}} \> \\[-.75\baselineskip]
    \>\> \pcind $NVPCRS \gets NVPCRS$ \\
    \>\> \pcind $\quad \cup \, \langle idx, \hash(0\ldots0 \, \concat \, IV) \rangle$
    }
    \end{bbrbox}

    \node[draw, dashed, line width=\lightrulewidth, above=3.1cm of D.north east, anchor=east] (orc1) {\pseudocode[mode=text, codesize=\large, linenumbering]{
        $IV \gets 0\ldots0$\\
        $\textbf{if } \mathcal{B}_{NV} = \false \textbf{ then }$\\ \pcind $REQ_{add} \gets \set{\mathcal{B}_{NV}, idx}$\\
        \pcind $mPCR^\mathcal{X} \gets mPCR^\mathcal{X} \cup \langle idx, \hash(0\ldots0\, \concat \, IV) \rangle$\\
        $\textbf{else }$\\
        \pcind $h_{pol} \gets \hash(\hash(\hash(0\ldots0 \, \concat \, CC_{PolicyAuthorize} \, \concat \, \textsf{name}(EK^{\orc})))$ \pcskipln\\
        \pcind $\quad \concat \, CC_{PolicyCommandCode} \, \concat \, CC_{NV\_UndefineSpaceSpecial})$\\
        \pcind $REQ_{add} \gets \set{idx, \mathsf{TPL}(idx), h_{pol}, IV, \mathcal{B}_{NV}}$
    }};
    \node[anchor=north east, xshift=-0.1cm, yshift=-0.1cm] at (orc1.north east) {\large{$\bm{\orc}$}};
    \draw[->, line width=\heavyrulewidth] ([xshift=-2.5cm]orc1.south east)--([xshift=-2.5cm, yshift=0.1cm]D.north east) node [midway, left] {\large $REQ_{add}$};

    \draw[->, line width=\heavyrulewidth] ([xshift=-2.4cm]orc1.west)--([xshift=-0.15cm]orc1.west) node [midway, above] {\large $\mathcal{B}_{NV}, idx, \mathcal{X}$};

    \node[draw, dashed, line width=\lightrulewidth, below=3cm of D.south east, anchor=east] (orc2) {\pseudocode[mode=text, codesize=\large, linenumbering, lnstart=3]{
        $\verify(certInfo.magic = \mathsf{TPM\_GENERATED})\quad\quad\quad$\\
        $\verify(certInfo.nvContents = \hash(0\ldots0 \, \concat \, IV))$\\
        $\verify(certInfo.objName = \textsf{name}(\textsf{TPL}(idx) \cup \set{\mathsf{WRITTEN}, idx,  h_{pol}})$ \\
        $\enc(\sig^{EK^\mathcal{X}_\sk}_{certInfo}, EK^\mathcal{X}_\pk) = certInfo)$\\
        $mNVPCR^\mathcal{X} \gets mNVPCR^\mathcal{X} \cup \langle idx, \hash(0\ldots0 \, \concat \, IV), certInfo.objName \rangle$
    }};
    \node[anchor=north east, xshift=-0.1cm, yshift=-0.1cm] at (orc2.north east) {\large{$\bm{\orc}$}};
    \draw[->, line width=\heavyrulewidth] ([xshift=-2.5cm]D.south east)--([xshift=-2.5cm, yshift=0.1cm]orc2.north east) node [midway, left] {\large $\set{\sig^{EK^\mathcal{X}_\sk}_{certInfo}, certInfo}$};
    
    \end{bbrenv}
    \end{adjustbox}
    \caption{Attaching a normal or NV-based PCR}
    \label{fig:createNvPcr}
\end{figure}\normalsize

\begin{algorithm}[htbp]
    \DontPrintSemicolon
    \SetKwInOut{Input}{Input}
    \SetKwInOut{Output}{Output}
    \Input{$n, idx, \mathcal{H}_{EK}, mNVPCR$}
    \Output{$\set{idx, h_{cp}, \sig^{EK^{\orc}_\sk}_{aHash}, h_{pol}, \hash(h_{pol}), \sig^{EK^{\orc}_\sk}_{\hash(h_{pol})}}$}
    $h_{pol} \gets \hash(\hash(0\ldots0 \, \concat \, CC_{PolicySigned} \, \concat \, \textsf{name}(\mathcal{H}_{EK})))$\;
    $\sig^{EK^{\orc}_\sk}_{\hash(h_{pol})} \gets tpm.\sign(\hash(h_{pol}), \mathcal{H}_{EK}))$\;
    $h_{cp} \gets \emptyset$\;
    $\forall \langle \mathcal{H}, h, \mathsf{name}(\mathcal{H}) \rangle \in mNVPCR:$\;
    \Indp$\textbf{if } \mathcal{H} = idx \textbf{ then}$\;
    \Indp$h_{cp} \gets \hash(CC_{NV\_UndefineSpaceSpecial}$ $\quad\concat \, \mathsf{name}(\mathcal{H}) \, \concat \, \mathcal{H}_{PPS})$\;
    \Indm\Indm$aHash \gets \hash(n \, \concat \, 0 \, \concat \, h_{cp})$\;
    $\sig^{EK^{\orc}_\sk}_{aHash} \gets tpm.\sign(aHash, \mathcal{H}_{EK})$\;
    \textbf{return} $idx, h_{cp}, \sig^{EK^{\orc}_\sk}_{aHash}, h_{pol}, \hash(h_{pol}), \sig^{EK^{\orc}_\sk}_{\hash(h_{pol})}$\;
    \caption{Authorizing NV index deletion}
    \label{alg:authorizedNvDeletion}
\end{algorithm}

\begin{figure}[htbp]
    \centering
    \begin{adjustbox}{
    , width=1\columnwidth}
    \begin{bbrenv}{E}
    
    \begin{bbrbox}[abovesep=0pt]
    \pseudocode[mode=text, colsep=0pt, colspace=0pt, jot=0mm]{
    $\bm{\mathcal{TC}} : \textsc{\textbf{vTpm}}$ \>\> $\bm{\mathcal{X}: \vf}$\\
    $PPS$ \>\> $\mathcal{H}_{EK^{\orc}_\pk}, PCRS, NVPCRS$ \\ [][\bottomrule] \\[-0.5\baselineskip]
    \>\> $\textbf{if } \mathcal{B}_{NV} = \false \textbf{ then}$ \\
    \>\> \pcind $PCRS \gets PCRS \setminus idx$ \\
    \>\> $\textbf{else }$ \\
    \> \sendmessage{<-}{style={line width=\lightrulewidth}, top=$\text{Session type}: \mathsf{POLICY}$} \> \\[-2.2\baselineskip]
    \xdash[17em] \>\> \\[-0.5\baselineskip]
    $\bm{\mathsf{TPM2\_StartAuthSession}}$ \>\> \\
    $\mathcal{H}_\mathsf{ps} \gets \mathsf{session}(\mathsf{POLICY})$ \>\> \\
    $\mathsf{session}(\mathcal{H}_\mathsf{ps}) . h_{pol} \gets 0\ldots0$ \>\> \\
    $n \sample \bin^t$ \>\> \\
    \> \sendmessage{->}{style={line width=\lightrulewidth}, top={$\mathcal{H}_\mathsf{ps}, n$}} \> \\
    \> \sendmessage{<-}{style={line width=\lightrulewidth}, top={$\hash(h_{pol}), \sig^{EK^{\orc}_\sk}_{\hash(h_{pol})}, \mathcal{H}_{EK^{\orc}_\pk}$}} \> \\[-2.2\baselineskip]
    \xdash[17.5em] \>\> \\[-0.5\baselineskip]
    $\bm{\mathsf{TPM2\_VerifySignature}}$ \>\> \\
    $\verify(\enc(\sig^{EK^{\orc}_\sk}_{\hash(h_{pol})}, \mathcal{H}_{EK^{\orc}_\pk}) = \hash(h_{pol}))$ \>\> \\
    $t \gets \mathsf{HMAC}(\mathsf{proof}(\mathcal{H}_{EK^{\orc}_\pk}), (\mathsf{VERIFIED}$ \>\> \\
    $\quad \concat \, \hash(h_{pol}) \, \concat \, \mathsf{name}(EK^{\orc}_\pk)))$ \>\> \\
    \> \sendmessage{->}{style={line width=\lightrulewidth}, top={$t$}} \> \\
    \> \sendmessage{<-}{style={line width=\lightrulewidth}, top={$\sig^{EK^{\orc}_\sk}_{aHash}, h_{cp}, n, \mathcal{H}_{EK^{\orc}_\pk}, \mathcal{H}_\mathsf{ps}$}} \> \\[-2.2\baselineskip]
    \xdash[17.5em] \>\> \\[-0.5\baselineskip]
    $\bm{\mathsf{TPM2\_PolicySigned}}$ \>\> \\
    $aHash \gets \hash(n \, \concat \, 0 \, \concat \, h_{cp})$ \>\> \\
    $\verify(\enc(\sig^{EK^{\orc}_\sk}_{aHash}, \mathcal{H}_{EK^{\orc}_\pk}) = aHash)$ \>\> \\
    $\mathsf{session}(\mathcal{H}_\mathsf{ps}) . h_{pol} \gets \hash(\mathsf{session}(\mathcal{H}_\mathsf{ps}) . h_{pol}$ \>\> \\
    $\quad \concat \, CC_{PolicySigned}$ \>\> \\
    $\quad \concat \, \mathsf{name}(EK^{\orc}_\pk))$ \>\> \\
    $\mathsf{session}(\mathcal{H}_\mathsf{ps}) . h_{pol} \gets \hash(\mathsf{session}(\mathcal{H}_\mathsf{ps}) . h_{pol})$ \>\> \\
    $\mathsf{session}(\mathcal{H}_\mathsf{ps}) . cpHash \gets h_{cp}$ \>\> \\
    \> \sendmessage{<-}{style={line width=\lightrulewidth}, top={$h_{pol}, t, \mathsf{name}(EK^{\orc}_\pk), \mathcal{H}_\mathsf{ps}$}} \> \\[-2.2\baselineskip]
    \xdash[17em] \>\> \\[-0.5\baselineskip]
    $\bm{\mathsf{TPM2\_PolicyAuthorize}}$ \>\> \\
    $\verify(\mathsf{session}(\mathcal{H}_\mathsf{ps}) . h_{pol} = h_{pol})$ \>\> \\
    $t \gets \mathsf{HMAC}(\mathsf{proof}(t), (\mathsf{VERIFIED}$ \>\> \\
    $\quad \concat \, \hash(h_{pol}) \, \concat \, \mathsf{name}(EK^{\orc}_\pk)))$ \>\> \\
    $\verify(t = t)$ \>\> \\
    $\mathsf{session}(\mathcal{H}_\mathsf{ps}) . h_{pol} \gets 0\ldots0$ \>\> \\
    $\mathsf{session}(\mathcal{H}_\mathsf{ps}) . h_{pol} \gets \hash(\mathsf{session}(\mathcal{H}_\mathsf{ps}) . h_{pol}$ \>\> \\
    $\quad \concat \, CC_{PolicyAuthorize} \, \concat \, \mathsf{name}(EK^{\orc}_\pk))$ \>\> \\
    \>\> \\
    \> \sendmessage{<-}{style={line width=\lightrulewidth}, top={$CC_{NV\_UndefineSpaceSpecial}, \mathcal{H}_\mathsf{ps}$}} \> \\[-2.2\baselineskip]
    \xdash[17em] \>\> \\[-0.5\baselineskip]
    $\bm{\mathsf{TPM2\_PolicyCommandCode}}$ \>\> \\
    $\mathsf{session}(\mathcal{H}_\mathsf{ps}) . CC \gets$ \>\> \\ 
    $\quad CC_{NV\_UndefineSpaceSpecial}$ \>\> \\
    $\mathsf{session}(\mathcal{H}_\mathsf{ps}) . h_{pol} \gets \hash(\mathsf{session}(\mathcal{H}_\mathsf{ps}) . h_{pol}$ \>\> \\
    $\quad \concat \, CC_{PolicyCommandCode}$ \>\> \\
    $\quad \concat \, CC_{NV\_UndefineSpaceSpecial})$ \>\> \\
    \> \sendmessage{<-}{style={line width=\lightrulewidth}, top={$idx, \mathcal{H}_{PPS}, \mathcal{H}_\mathsf{ps}$}} \> \\[-2.2\baselineskip]
    \xdash[17em] \>\> \\[-0.5\baselineskip]
    $\bm{\mathsf{TPM2\_NV\_UndefineSpaceSpecial}}$ \>\> \\
    $\verify(\mathsf{authPol}(idx = \mathsf{session}(\mathcal{H}_\mathsf{ps}) . h_{pol})$ \>\> \\
    $\verify(\mathsf{session}(\mathcal{H}_\mathsf{ps}).CC =$ \>\> \\
    $\quad CC_{NV\_UndefineSpaceSpecial})$ \>\> \\
    $\verify(\mathsf{session}(\mathcal{H}_\mathsf{ps}).cpHash = \hash($ \>\> \\
    $\quad CC_{NV\_UndefineSpaceSpecial}$ \>\> \\
    $\quad \concat \, \mathsf{name}(idx) \, \concat \, \mathcal{H}_{PPS}))$ \>\> \\
    $\mathsf{NVDestroy}(idx) \land \mathsf{Destroy}(\mathcal{H}_\mathsf{ps})$ \>\> \\
    \>\> \pcind $\forall \langle \mathcal{H}_i, h_i \rangle \in NVPCRS :$ \\
    \>\> \pcind\pcind $\textbf{if } \mathcal{H}_i = idx \textbf{ then }$ \\
    \>\> \pcind\pcind $\quad NVPCRS \gets $ \\
    \>\> \pcind\pcind $\quad\quad NVPCRS \setminus \langle \mathcal{H}_i, h_i \rangle$
    }
    \end{bbrbox}

    \node[draw, dashed, line width=\lightrulewidth, above=3.2cm of E.north east, anchor=east] (orc1) {\pseudocode[mode=text, codesize=\large, linenumbering]{
        $\textbf{if } \mathcal{B}_{NV} = \false \textbf{ then }\quad\quad\quad$\\
        \pcind $REQ_{delete} \gets \set{idx}$\\
        \pcind $\forall \langle idx^\prime, h \rangle \in mPCR^\mathcal{X}: \textbf{if } idx^\prime = idx \textbf{ then }$\\
        \pcind\pcind $mPCR^\mathcal{X} \gets mPCR^\mathcal{X} \setminus \langle idx^\prime, h \rangle$\\
        $\textbf{else }$\\ 
        \pcind $REQ_{delete} \gets \text{Algorithm~\ref{alg:authorizedNvDeletion}}(n, idx, \mathcal{H}_{EK^{\orc}}, mNVPCR^{\vf})$\\
        \pcind $\forall \langle \mathcal{H}, h, \mathsf{name}(\mathcal{H}) \rangle \in mNVPCR^\mathcal{X}: \textbf{if } \mathcal{H} = idx \textbf{ then }$\\
        \pcind\pcind $mNVPCR^\mathcal{X} \gets mNVPCR^\mathcal{X} \setminus \langle \mathcal{H}, h, \mathsf{name}(\mathcal{H}) \rangle$
    }};
    \node[anchor=north east, xshift=-0.1cm, yshift=-0.1cm] at (orc1.north east) {\large{$\bm{\orc}$}};
    \draw[->, line width=\heavyrulewidth] ([xshift=-.5cm]orc1.south east)--([xshift=-.5cm, yshift=0.1cm]E.north east) node [midway, left] {\large $REQ_{delete} \cup \mathcal{B}_{NV}$};
    \draw[->, line width=\heavyrulewidth] ([xshift=-5cm, yshift=0.1cm]E.north east)--([xshift=-5cm]orc1.south east) node [midway, left] {\large $n$};

    \draw[->, line width=\heavyrulewidth] ([xshift=-2.4cm]orc1.west)--([xshift=-0.15cm]orc1.west) node [midway, above] {\large $\mathcal{B}_{NV}, idx, \mathcal{X}$};
    
    \end{bbrenv}
    \end{adjustbox}
    \caption{Detatching a normal or NV-based PCR}
    \label{fig:pcrRevocation}
\end{figure}

\subsection{Performance Evaluation}
\label{appendix:performanceEvaluation}

\subsubsection{Environmental Setup}
\label{subsec:environmentalSetup}

We implemented our protocols in C++ with IBM's TPM Software Stack (TSS) v1.6.0~\cite{ibmTsstpm} and OpenSSL v1.1.1i, compiled using the GNU GCC compiler. We considered only elliptic curve (EC) keys and used SHA256 as $\mathcal{H}$. We tested the protocols on two platforms: (\emph{\textbf{P1}}) a computer running the Windows 10 OS, equipped with a 3.6 GHz AMD Ryzen 7 3700X CPU, and running IBM's SW TPM v1637~\cite{ibmTsstpm}, and (\emph{\textbf{P2}}) a Raspberry Pi 4 Model B with an 1.5Ghz ARM Cortex-A72 CPU running the Raspbian (buster) OS with an TPM 2.0 compliant OPTIGA HW TPM SLB9670.


\subsubsection{Timing Tests}
\label{appendix:timings}

Table~\ref{tab:appendixTimings} shows the mean (M) and standard deviations (SD) after running each protocol 50 times on each platform (Section~\ref{subsec:environmentalSetup}). For each protocol, we show: (first row) how long it takes to complete the protocol (i.e., with preparation) and (next rows) how much time is allocated to each of the TPM commands. The timings are produced using C++11's chrono library's system clock; each timing statistic includes time spent on the program code, TSS processing, the TPM's internal processing, and any Low Pin Count (LPC) bus delay (for \emph{\textbf{P2}}). Note that verification of AK creation, NVPCR creation, and the signed challenge are omitted since they do not require interaction with the TPM and take little time, i.e., $\approx$ 0.5ms and 2.4ms on average for \emph{\textbf{P1}} and \emph{\textbf{P2}}, respectively.

Although a security-centered HW-TPM is a bottleneck when it comes to efficiency, it provides security guarantees that a SW-TPM cannot. Note that the most time-consuming protocols (i.e., AK creation, configuration updates, and NVPCR deletion) are executed intermittently between $\orc$ and $\vf$; thus, they have a negligible impact on the $\mathcal{SG}$. The attestation (ORA), which $\vf$s run among themselves, takes a $\vf$ (with two active PCRs, one normal and one NV-based), $<$0.4s to complete on a HW-TPM and $\approx$ 10ms with a SW-TPM. Note, however, that the efficiency of ORA depends on how many PCRs are attached to the $\vf$.

\begin{table}[htbp]
    \caption{Timings (in ms) for each platform setup.}
    \begin{center}
        \begin{tabular*}{\columnwidth}{@{\extracolsep{\fill}} lrrrr}
            \toprule
            Protocol & M (\emph{\textbf{P1}}) & $\pm$SD & M (\emph{\textbf{P2}}) & $\pm$SD\\
            \midrule
\textbf{AK creation} $\bm{(\vf)}$ & 96.20 & 1.00 & 543.23 & 6.66\\
\hspace{0.15cm}TPM2\_Create & 2.76 & 0.43 & 202.97 & 0.81\\
\hspace{0.15cm}TPM2\_Load & 2.98 & 0.42 & 56.61 & 1.79\\
\hspace{0.15cm}TPM2\_CertifyCreation & 1.12 & 0.33 & 146.35 & 2.29\\
\hspace{0.15cm}TPM2\_EvictControl & 3.18 & 0.52 & 97.87 & 1.77\\
\hspace{0.15cm}TPM2\_FlushContext & 5.44 & 0.54 & 37.11 & 1.52\\
\textbf{Measurement update} $\bm{(\vf)}$ & 9.63 & 4.55 & 392.58 & 3.33\\
\hspace{0.15cm}TPM2\_VerifySignature & 0.93 & 0.26 & 116.12 & 0.71\\
\hspace{0.15cm}TPM2\_StartAuthSession & 1.52 & 0.52 & 31.65 & 0.63\\
\hspace{0.15cm}TPM2\_NV\_Extend & 4.82 & 0.65 & 82.68 & 1.21\\
\hspace{0.15cm}TPM2\_PCR\_Extend & 4.84 & 5.47 & 79.37 & 1.13\\
\hspace{0.15cm}TPM2\_GetSessionAuditDigest & 1.14 & 0.35 & 128.23 & 0.85\\
\textbf{ORA} $\bm{(\prover)}$ & 9.84 & 8.34 & 386.68 & 2.96\\
\hspace{0.15cm}TPM2\_StartAuthSession & 1.52 & 0.52 & 31.65 & 0.63\\
\hspace{0.15cm}TPM2\_PolicyNV & 0.24 & 0.43 & 61.96 & 0.63\\
\hspace{0.15cm}TPM2\_PolicyPCR & 0.18 & 0.38 & 59.35 & 0.51\\
\hspace{0.15cm}TPM2\_PolicyAuthorize & 0.18 & 0.38 & 69.13 & 0.58\\
\hspace{0.15cm}TPM2\_Sign & 5.78 & 6.77 & 129.86 & 1.39\\
\textbf{Attaching a NVPCR} $\bm{(\vf)}$ & 9.14 & 0.60 & 113.16 & 1.60\\
\hspace{0.15cm}TPM2\_NV\_DefineSpace & 2.52 & 0.50 & 26.67 & 0.81\\
\hspace{0.15cm}TPM2\_NV\_Extend & 4.82 & 0.65 & 82.68 & 1.21\\
\hspace{0.15cm}TPM2\_NV\_Certify & 1.20 & 0.40 & 75.18 & 0.61\\
\textbf{Detatching a NVPCR} $\bm{(\vf)}$ & 8.98 & 0.62 & 524.93 & 2.54\\
\hspace{0.15cm}TPM2\_StartAuthSession & 1.52 & 0.52 & 31.65 & 0.63\\
\hspace{0.15cm}TPM2\_VerifySignature & 0.93 & 0.26 & 116.12 & 0.71\\
\hspace{0.15cm}TPM2\_PolicySigned & 0.90 & 0.30 & 163.50 & 0.82\\
\hspace{0.15cm}TPM2\_PolicyAuthorize & 0.18 & 0.38 & 69.13 & 0.58\\
\hspace{0.15cm}TPM2\_PolicyCommandCode & 0.16 & 0.37 & 58.40 & 0.83\\
\hspace{0.15cm}TPM2\_NV\_UndefineSpaceSpecial & 6.18 & 0.52 & 62.60 & 0.95\\
            \bottomrule
        \end{tabular*}
        \label{tab:appendixTimings}
    \end{center}
\end{table}\normalsize

\end{document}